\documentclass[
reprint,
 amsmath,
 amssymb,
 aip
 ]{revtex4-2}

 \usepackage{graphicx}
\usepackage{graphicx,xcolor,hyperref}
\usepackage{amsmath,amssymb,bm}
\usepackage{mathtools}
\usepackage{comment}
\graphicspath{{./figures/}}

\newcommand{\bx}{\bm{x}}
\newcommand{\bX}{\bm{X}}
\newcommand{\bA}{\bm{A}}
\newcommand{\bB}{\bm{B}}
\newcommand{\bR}{\bm{R}}
\newcommand{\by}{\bm{y}}
\newcommand{\bz}{\bm{z}}
\newcommand{\bU}{\bm{U}}

\newcommand{\bSig}{\bm{\Sigma}}
\newcommand{\bV}{\bm{V}}
\newcommand{\bPhi}{\bm{\Phi}}
\newcommand{\bphi}{\bm{\phi}}
\newcommand{\bLam}{\bm{\Lambda}}

\begin{document}
\title{Dynamic mode decomposition for detecting oscillatory transient activity via sparsity and smoothness regularization}

\author{Yutaro Tanaka}
\affiliation{Department of Systems and Control Engineering, Institute of Science Tokyo, Tokyo 152-8552, Japan}
\email{tanaka.y.8c8a@m.isct.ac.jp}
\author{Hiroya Nakao}
\affiliation{Department of Systems and Control Engineering, Institute of Science Tokyo, Tokyo 152-8552, Japan}
\affiliation{Research Center for Autonomous Systems Materialogy, Institute of Science Tokyo, Yokohama 226-8501, Japan}

%%%%%%%%%%%%%%%%%%%%%%%%%%%%%%%%%%%%%%%%%%%%%%%%%%%%%%%%
%%% Abstract %%%%%%%%%%%%%%%%%%%%%%%%%%%%%%%%%%%%%%%%%%%
%%%%%%%%%%%%%%%%%%%%%%%%%%%%%%%%%%%%%%%%%%%%%%%%%%%%%%%%
\begin{abstract}
Dynamic mode decomposition (DMD) is a data-driven modal decomposition technique that extracts coherent spatiotemporal structures from high-dimensional time-series data. 
By decomposing the dynamics into a set of modes, each associated with a single frequency and a growth rate, DMD enables a natural modal decomposition and dimensionality reduction of complex dynamical systems.
However, when DMD is applied to transient dynamics, even if a large number of modes are used, 
it remains difficult to interpret how these modes contribute to the transient behavior.
In this study, we propose a simple extension of DMD that facilitates extraction of oscillatory transient activity by introducing time-varying amplitudes for the DMD modes based on sparsity and smoothness regularization. 
This approach enables the identification of dynamically significant modes and extraction of their transient activities, providing a more interpretable representation of non-steady dynamics. 
We illustrate the validity of the proposed method using a simple example and then apply it to fluid flow data of a laminar airfoil wake exhibiting transient behavior. 
We demonstrate that it can capture the temporal structure of mode activations that are not accessible with the standard DMD method.
\end{abstract}
%%%%%%%%%%%%%%%%%%%%%%%%%%%%%%%%%%%%%%%%%%%%%%%%%%%%%%%%
\maketitle
{\bf
Complex dynamical phenomena are observed in various fields of science and engineering.
As explicit mathematical models are often unavailable, data-driven methods for extracting essential spatiotemporal structures from time series are gaining importance.
Dynamic mode decomposition (DMD) is a convenient and widely used method for this purpose, but the standard DMD fails to characterize the underlying dynamics when applied to systems exhibiting transient behavior.
In this work, we address this issue by introducing a simple extension to DMD that can extract oscillatory transient activity of the DMD modes.
The proposed framework is general and can be integrated with other DMD-based methods for diverse applications. 
}
\section{Introduction}
%%%%%%%%%%%%%%%%%%%%%%%%%%%%%%%%%%%%%%%%%%%%%%%%%%%%%%%%
%%% Introduction %%%%%%%%%%%%%%%%%%%%%%%%%%%%%%%%%%%%%%%
%%%%%%%%%%%%%%%%%%%%%%%%%%%%%%%%%%%%%%%%%%%%%%%%%%%%%%%%

In many scientific and engineering applications, the underlying governing equations that describe a system's dynamics are often not explicitly known.
Instead, we have access only to time-series data of the system's variables, obtained either through simulations or experimental measurements.
These time-series data are typically assumed to be generated by a dynamical system, 
\begin{align}
    \frac{d\bx(t)}{dt} = \bm{f}(\bx(t),t), \label{Eq:Dynamical System}
\end{align}
where $\bx\in\mathbb{R}^{N}$ denotes the system state in an $N$-dimensional space and $\bm{f}:\mathbb{R}^{N} \times \mathbb{R} \to\mathbb{R}^{N}$ is an unknown vector-valued function describing the dynamics of the system.
Understanding the qualitative and quantitative behavior of such systems from observed data without explicit knowledge of $\bm{f}$ remains a fundamental challenge~\cite{Brunton Book,ResDMD}.

Recent advances in sensing technologies, data acquisition systems, and high-performance computing have enabled the collection of increasingly rich, high-dimensional, and temporally resolved data~\cite{Brunton Book}. 
These developments have led to the growing importance of data-driven modeling techniques, which aim to approximate, predict, and understand dynamical systems directly from data.
Among those, dynamic mode decomposition (DMD) is a particularly convenient and widely used approach~\cite{DMD,DMD Review1}.
It provides a purely data-driven framework for analyzing spatiotemporal patterns in complex systems by decomposing the data into a set of spatial modes, each associated with  a specific frequency and a growth rate.

The key advantage of DMD lies in its ability to produce a linear decomposition and reduced-order representation of potentially nonlinear systems that are interpretable and computationally efficient.
Moreover, DMD is closely related to Koopman operator theory~\cite{KOT1,KOT2}, which provides a linear representation of nonlinear dynamical systems in an infinite-dimensional function space. 
This connection has further solidified the theoretical foundation of DMD~\cite{EDMD,Koopman analysis flow1} and expanded its applicability across a wide range of fields, including fluid dynamics~\cite{DMD}, nonlinear oscillations~\cite{Kato,Takata}, cellular automata~\cite{cellular automata1,cellular automata2}, neuroscience~\cite{Neuroscience}, robotics~\cite{robot1,robot2}, epidemiology~\cite{epidemiology}, and financial modeling~\cite{financial trading}.

Despite its strength, the standard formulation of DMD is inherently limited by the  assumption of stationarity.
However, many real world systems exhibit non-steady behavior, especially transient processes, where the dominant patterns of the system change over time; for example, fluid undergoing transition between different flows, neuronal activity during task switching, or financial markets responding to external shocks.
In the standard DMD, the temporal behavior of each mode is characterized solely by an eigenvalue and a constant amplitude.
As a result, when DMD is applied to transient dynamics, even if a large number of modes and associated eigenvalues are used, the temporal activation of each mode and its contribution to the dynamics cannot be captured.
However, if such temporal structures are not taken into account, interpretation and predictions of the system behavior based on the extracted modes may lack reliability.

To address this limitation, several extensions of DMD have been proposed. 
One of the earliest 
approaches is the multiresolution DMD (mrDMD)~\cite{mrDMD}, which decomposes the data across multiple temporal scales. 
Instead of applying DMD to the entire dataset, mrDMD performs a hierarchical, recursive decomposition of the time series, enabling the resolution of dynamics that are localized in both time and frequency. 
Various windowed DMD methods have also been developed, which apply DMD to short, time-localized segments of the data to track the time evolution of the underlying system~\cite{Window DMD1,Window DMD2}.

More broadly, Koopman spectral analysis or DMD for non-stationary dynamical systems has been investigated in various studies.
Proctor {\em et al}.~\cite{DMDc} introduced DMD with control (DMDc), which accounts for external inputs in the identification of dynamic modes. 
Mezi\'{c} and Surana~\cite{Koopman periodic} extended the Koopman operator framework to non-autonomous system with periodic or quasi-periodic time dependence by introducing time-parametrized operator family and time-dependent spectral components.
Mace\v{s}i\'c {\em et al}.~\cite{Koopman Nonauto} proposed an algorithm for computing the spectrum of the Koopman operator in non-autonomous systems by applying the fundamental matrix of linear time-varying systems and Arnoldi-like methods to local time-windowed data. 
Redman {\em et al}.~\cite{Koopman episodic memory} introduced an episodic memory approach that records spectral data from localized Koopman operator approximations and reuses the stored information for forecasting.

However, approaches like mrDMD and windowed DMD rely on a deterministic, predefined partitioning of the data. 
Their performance is sensitive to the choice of window sizes and segmentation locations, which may not align with the natural time scales of the underlying dynamics and can hinder accurate extraction of the modes from systems with complex transient behaviors. This motivates the need for an adaptive, data-driven approach that can capture non-stationary dynamics without relying on rigid segmentation.

In this paper, we propose a simple extension to the standard DMD framework that introduces time-dependent mode amplitudes based on sparsity and smoothness regularization, which is useful in extracting oscillatory transients from data.
Unlike mrDMD or windowed DMD approaches, which partition the data into predetermined temporal segments and analyze each segment separately, our method applies DMD to the entire time-series dataset as a whole, focusing on the modes and frequencies extracted from the entire data structure.
By treating mode amplitudes as smoothly varying functions over the full time horizon and enforcing sparsity in a data-driven manner, 
we can avoid ad hoc windowing choices and automatically detect when specific modes become active or inactive. 
We validate the proposed method on flow datasets that exhibit transient dynamics and demonstrate that our approach provides a more  faithful and interpretable characterization of the system compared to the standard DMD.

\section{Dynamic mode decomposition}
%%%%%%%%%%%%%%%%%%%%%%%%%%%%%%%%%%%%%%%%%%%%%%%%%%%%%%%%
%%% Dynamic mode decomposition %%%%%%%%%%%%%%%%%%%%%%%%%
%%%%%%%%%%%%%%%%%%%%%%%%%%%%%%%%%%%%%%%%%%%%%%%%%%%%%%%%

In this section, we briefly review DMD. 
It was initially proposed in fluid mechanics as a method to extract spatiotemporal coherent structures~\cite{DMD}.
Owing to its conceptual simplicity and practical utility, DMD has been extended in numerous directions, including DMDc~\cite{DMDc}, mrDMD~\cite{mrDMD}, sparsity-promoting DMD~\cite{spDMD}, recursive DMD~\cite{RDMD}, and physics-informed DMD~\cite{piDMD}.
DMD methods to improve robustness to noise~\cite{optDMD,noiseDMD} have also been developed, and 
explicit connection of DMD with the Koopman operator theory has also been established~\cite{ResDMD,EDMD,Koopman analysis flow1,KDMD}.
At present, the most used definition and algorithm of DMD is the exact DMD~\cite{Exact DMD}, which provides a closed-form solution based on the Moore--Penrose pseudo-inverse.

We assume that the system is governed by Eq.~\eqref{Eq:Dynamical System} and that we can obtain discrete time snapshots of the system, i.e., $M$ snapshots with a time interval $\Delta t$,
\begin{align*}
    \{\bx_{0}, \bx_{1},...,\bx_{M}\},
\end{align*}
where each $\bx_{m}=\bx(m\Delta t)$ is an $N$-dimensional vector.
These snapshots are arranged into two data matrices,
\begin{align}
    \begin{split}
    \bX_{-}&=
    \begin{bmatrix}
        \bx_{0} & \bx_{1} & \cdots & \bx_{M-1}
    \end{bmatrix}\in {\mathbb{R}}^{N\times M},\\
    \bX_{+}&=
    \begin{bmatrix}
        \bx_{1} & \bx_{2} & \cdots & \bx_{M}
    \end{bmatrix}\in {\mathbb{R}}^{N\times M}. \label{Eq:Data Matrix}
\end{split}
\end{align}

The exact DMD algorithm~\cite{Exact DMD, DMD Book} seeks a linear operator $\bA \in {\mathbb R}^{N \times N}$ that gives the best linear approximation to the evolution of the snapshot measurements forward in time, 
\begin{align}
    \bx_{m+1}\approx\bA\bx_{m},\ \ m=\{0,1,...,M-1\} \label{Eq:Linear Dynamics}.
\end{align}
Expressed using the data matrices in Eq.~\eqref{Eq:Data Matrix}, the linear dynamics in Eq.~\eqref{Eq:Linear Dynamics} can be written as
\begin{align*}
    \bX_{+} \approx \bA\bX_{-}.
\end{align*}
Such a linear operator $\bA$ that minimizes the approximation error is given by
\begin{align}
        \bA = \underset{\bB\in {\mathbb{R}}^{N\times N}}{\mathrm{argmin}}||\bX_{+}-\bB\bX_{-}||_{F}=\bX_{+}\bX_{-}^{\dag},\label{Eq:Opt Problem}
\end{align}
where $||\cdot||_{F}$ is the Frobenius norm and $\dag$ denotes the Moore--Penrose pseudo-inverse.

In many systems like fluid flows, the system's dimension is very large, i.e., $N \gg M$, and the matrix $\bA$ cannot be treated directly. 
Therefore, rather than using Eq.~\eqref{Eq:Opt Problem} explicitly, the DMD algorithm forms a smaller matrix $\tilde{\bA} \in {\mathbb R}^{r \times r}$ ($r \leq M$), a reduced-order representation of $\bA$.
First, the singular value decomposition of $\bX_{-}$ is computed, i.e., $\bX_{-}\approx\bm{U}\bSig\bV^{*}$, where $*$ represents the Hermite conjugate, 
$\bU\in\mathbb{R}^{N\times r}$ and $\bV\in\mathbb{R}^{M\times r}$ contain the singular vectors, and $\bSig\in\mathbb{R}^{r\times r}$ is a diagonal matrix with the singular values.
The truncation rank $r\leq M$ of $\bX_{-}$ is determined based on the decay of the singular values of $\bX_{-}$.
From Eq.~\eqref{Eq:Opt Problem},
$\bA$ is projected onto the leading $r$ SVD modes of $\bX_{-}$ as
\begin{align*}
    \bA = \bX_{+}\bX_{-}^{\dag} \approx \bX_{+}\bV\bSig^{-1}\bU^{*},
\end{align*}
and the reduced matrix $\tilde{\bA}$ is obtained as
\begin{align*}
    \widetilde{\bA} = \bU^{*}\bA\bU = \bU^{*}\bX_{+}\bV\bSig^{-1}.
\end{align*}
The eigendecomposition $\tilde{\bA}$ is given by
\begin{align*}
    \widetilde{\bA}=\widetilde{\bPhi}\bLam\widetilde{\bPhi}^{-1},
\end{align*}
where $\bLam =\mbox{diag}(\lambda_1, ..., \lambda_r)$ is a diagonal matrix containing eigenvalues $\{ \lambda_{j} \}$ of $\widetilde{\bA}$, which are part of the eigenvalues of $\bA$, and $\widetilde{\bPhi}$ is a matrix whose columns consist of eigenvectors of $\widetilde{\bA}$.
The eigenvectors of the original matrix $\bA$ can be approximately reconstructed from
\begin{align*}
    \bPhi = \bX_{+}\bV\bSig^{-1}\widetilde{\bPhi},
\end{align*}
where $\bPhi = \begin{bmatrix}
        \bphi_{1} & \cdots & \bphi_{r}
    \end{bmatrix}$ is a matrix consisting of the approximate eigenvectors $\{ \bphi_{j} \}$ of $\bA$.

The system state at time step $m$ can be decomposed by the eigenvectors $\{ \bphi_{j} \}$ and eigenvalues $\{ \lambda_{j} \}$ of $\bA$ as
\begin{align}
\begin{split}
    &\bx_{m}\approx\bPhi\bLam^{m}\bz=\sum_{j=1}^{r}\lambda_{j}^{m}z_{j}\bm{\phi}_{j}\\
    &=\begin{bmatrix}
        \bphi_{1} & \cdots & \bphi_{r}
    \end{bmatrix}
    \begin{bmatrix}
         \lambda_{1} &        &             \\
                     & \ddots &             \\
                     &        & \lambda_{r} \\ 
    \end{bmatrix}^{m}
    \begin{bmatrix}
        z_{1} \\ \vdots \\ z_{r}
    \end{bmatrix}, \label{Eq:Decomposition}
\end{split}
\end{align}
where the vector $\bz$ contains the initial amplitude of each eigenvector, i.e., $\bz=\bPhi^{\dag}\bx_{0}$, and $\bphi_{j}$ is called the $j$th DMD mode.
The eigendecomposition in Eq.~\eqref{Eq:Decomposition} is intimately associated with the Koopman operator of the system~\cite{Koopman analysis flow1,Koopman analysis flow2,Koopman Review,Nakao2}.
From the above decomposition, we can observe that the DMD modes $\{\bphi_{j}\}$ characterize the spatial patterns and the corresponding eigenvalues $\{\lambda_{j}\}$ characterize the temporal dynamics (frequencies and growth rates) of the system.

Due to its ability to extract frequency-associated spatial modes enabling direct identification of temporal dynamics, DMD is now widely used in various fields, in particular, in fluid mechanics~\cite{Model Analysis of Fluid Flows1,Model Analysis of Fluid Flows2}. 
While the classical proper orthogonal decomposition (POD)~\cite{POD and DMD} yields energy-optimal modes, it does not directly provide frequency information, making DMD particularly suited for the spectral and temporal analyses.

\section{DMD for detecting active modes} \label{Sec:Method}
%%%%%%%%%%%%%%%%%%%%%%%%%%%%%%%%%%%%%%%%%%%%%%%%%%%%%%%%
%%% DMD for detecting active modes %%%%%%%%%%%%%%%%%%%%%
%%%%%%%%%%%%%%%%%%%%%%%%%%%%%%%%%%%%%%%%%%%%%%%%%%%%%%%%

In this section, we propose a method that can capture and analyze transient dynamics based on DMD, in particular, sparsity-promoting DMD~\cite{spDMD}. 
To achieve this goal, we design our proposed method to meet three key requirements:
\begin{enumerate}
    \item \emph{Switching ``on/off'' of each DMD mode.} At each time step, each DMD mode is adaptively switched ``on'' or ``off'' by imposing sparse regularization. 
    This allows us to extract which mode is contributing to the dynamics at which time step.
    \item \emph{Time-varying amplitudes.} We allow the amplitudes of DMD modes to vary with time.
     That is, we introduce $\bz_{m} = [ z_{1m}, ..., z_{rm} ]^{\top}$, which depends on the time step $m$, in place of $\bz = [ z_1, ..., z_r]^{\top}$ in Eq.~\eqref{Eq:Decomposition}.
     This enables us to capture the transient behavior of the system dynamics.
    \item \emph{Smoothness of amplitudes.} We impose a smoothness constraint on the amplitudes over time. 
    This allows us to capture the transient nature of the dynamics without introducing abrupt changes that could distort the representation.
\end{enumerate}

Considering the above three requirements, we start by modifying the decomposition in Eq.~\eqref{Eq:Decomposition} as 
\begin{align}
    \bx_{m} \approx \sum_{j=1}^{r}e^{im\omega_{j}}z_{jm}\bphi_{j}, 
    \label{Eq:Proposed Model}
\end{align}
where $z_{jm}$ represents the amplitude of the $j$th mode and $\omega_{j} = \mbox{arg}~(\lambda_{j} / | \lambda_{j}|)$ is its frequency, i.e., the argument of $\lambda_{j}$ satisfying $e^{i \omega_j} = \lambda_j / |\lambda_j|$.
Here, normalizing $\lambda_{j}$ by the absolute value $|\lambda_{j}|$ removes the effect of growth or decay, leaving only the phase information.
That is, we decompose the amplitude $\lambda_j^m z_j$ at time step $m$ into the phase part $e^{i m \omega_j}$ and the remaining amplitude $z_{jm}$.
We assume that the amplitude $z_{jm}$ varies smoothly, i.e., sufficiently more slowly than the phase $m\omega_{j}$ of the $j$th mode.
We also assume that the amplitudes $\{ z_{jm} \}$ are sparse and represent  dynamically active modes at each time step $m$.

Thus, we assume that any growth or decay that was supposed to be encoded in $|\lambda_{j}|^{m}$ in Eq.~\eqref{Eq:Decomposition}
is absorbed into the time-varying amplitude $z_{jm}$ in Eq.~\eqref{Eq:Proposed Model}. 
This modification keeps the information on growth or decay rates, even if the eigenvalues are not on the unit circle.
However, by further requiring that the amplitude $z_{jm}$ of the $j$th mode varies slowly with $m$ compared to its phase, we implicitly assume that the eigenvalues are relatively close to the unit circle.
Therefore, Eq.~\eqref{Eq:Proposed Model} might not be able to capture the dynamics of such a mode accurately if its growth or decay rate is very rapid.
Thus, our method focuses on capturing oscillatory transient modes with relatively slow decay rates. 

As described above, our method finds a smoothly varying $\bz_{m}$ characterizing the amplitudes of the DMD modes. 
We first obtain a set of pairs $\{(\bphi_{j},\lambda_{j})\}_{j=1}^{r}$ by using the exact DMD or some other DMD-type method~\cite{DMD Review1,DMD Book,DMD Review2}.
Then, we introduce a regularization framework based on sparsity and smoothness to extract the time-dependent amplitudes~\cite{spDMD,SINDy}.
Our method is formulated in the following optimization problem:
\begin{align}
    \begin{split}
    \bz_{m} & = \underset{\by\in\mathbb{C}^{r}}{\mathrm{argmin}}\left\| \bx_{m} - \bPhi\mathrm{diag}(\by)
    \begin{bmatrix}
        e^{i m \omega_{1}} \\ \vdots \\  e^{i m \omega_{r}}                               
    \end{bmatrix}
    \right\|_{2}^{2}\\
    & + \alpha_{2}\left\|\by-\bz_{m-1}\right\|_{2}^{2}
    + \alpha_{0} \|  {\bm y} \|_0,
    \ \ m=\{0,1,...,M\},
    \label{Eq:Opt Problem original}
\end{split}
\end{align}
where $\bPhi$ is a matrix consisting of the eigenvectors of $\bA$ obtained by DMD, $\| ... \|_{2}$ represents the $\ell_2$ norm, $\| ... \|_0$ is the $\ell_0$ norm, i.e., $\| \by \|_0$ is the number of non-zero elements of $\by =  [y_1, ..., y_r]^\top$, and $\alpha_2 \geq 0$ and $\alpha_0 \geq 0$ are regularization parameters.
For $m=0$, we assume $\bz_{-1}=\bm{0}$ and set $\alpha_{2}=0$.

In Eq.~\eqref{Eq:Opt Problem original}, the first term represents the reconstruction error between the observed data $\bx_{m}$ and the proposed decomposition of the form Eq.~\eqref{Eq:Proposed Model}, i.e., $\sum_{j=1}^{r}e^{im\omega_{j}} y_{j}\bphi_{j}$.
The second term is the $\ell_{2}$-penalty for the smoothness of the temporal amplitude of $\bz_{m}$, which encourages the solution to be close to the previous time step solution $\bz_{m-1}$,
where $\alpha_{2}$ controls the smoothness of the amplitude variations over time.
Note that the second term represents a discrete approximation to the squared time derivative of $\bz(t)$ scaled by $\Delta t^2$, i.e., $\alpha_2 \|\bz_m - \bz_{m-1}\|_2^2 \approx \alpha_2 \Delta t^2 \|\dot{\bz}\|^2$, which penalizes rapid variations in $\bz(t)$.
Therefore, if it is necessary to fix the physical smoothness strength at a prescribed level $\lambda$ independently of the sampling interval $\Delta t$ (i.e., to minimize $\lambda D_j$, where $D_j = \int|\dot{z}_j(t)|^2 dt$), the discrete parameter $\alpha_2$ should be set as $\alpha_2 \propto \lambda / \Delta t$.
The third $\ell_0$ term promotes the sparsity of the representation of $\bz_m$ by penalizing the number of non-zero elements in $\by$, where $\alpha_0$ controls the sparsity.

Equation~\eqref{Eq:Opt Problem original} gives a combinatorial optimization problem, which is difficult to solve.
Therefore, as in Ref.~\cite{spDMD} and other sparse regression methods, we instead consider the following $\ell_1$ optimization problem for $\bz_{m}$:

\begin{align}
    \begin{split}
    \bz_{m} & = \underset{\by\in\mathbb{C}^{r}}{\mathrm{argmin}}\left\| \bx_{m} - \bPhi\mathrm{diag}(\by)
    \begin{bmatrix}
        e^{i m \omega_{1}} \\ \vdots \\  e^{i m \omega_{r}}                               
    \end{bmatrix}
    \right\|_{2}^{2}\\
    & + \alpha_{2}\left\|\by-\bz_{m-1}\right\|_{2}^{2}
    + \alpha_{1}\left\|\by\right\|_{1},\ \ m=\{0,1,...,M\}, \label{Eq:Opt Problem For z}
\end{split}
\end{align}
where $\| ... \|_{1}$ represents the $\ell_1$ norm and $\alpha_{1}\geq 0$ is a regularization parameter that controls the sparsity of the solution.
Also, for high dimensional systems, i.e., $N\gg 1$, it is often difficult to solve the optimization problem in Eq.~\eqref{Eq:Opt Problem For z} directly.
In such cases, using the matrix $\bU$ of the singular value decomposition of $\bX_{-}$, we can consider the following approximate problem projected into a lower-dimensional subspace:
\begin{align}
    \begin{split}
    \bz_{m} & = \underset{\by\in\mathbb{C}^{r}}{\mathrm{argmin}}\left\| \bU^{*}\bx_{m} - \bU^{*}\bPhi\mathrm{diag}(\by)
    \begin{bmatrix}
        e^{i m \omega_{1}} \\ \vdots \\ e^{i m \omega_{r}}                               
    \end{bmatrix}
    \right\|_{2}^{2}\\
    & + \alpha_{2}\left\|\by-\bz_{m-1}\right\|_{2}^{2}
    + \alpha_{1}\left\|\by\right\|_{1},\ \ m=\{0,1,...,M\}.
\end{split}
\label{Eq:Opt Problem For z proj}
\end{align}

First, we determine the sparsity structure, i.e., which DMD modes to use at each time step $m$. We can solve Eq.~\eqref{Eq:Opt Problem For z} or~\eqref{Eq:Opt Problem For z proj} using the alternating direction method of multipliers (ADMM)~\cite{ADMM} for given $\alpha_2$ and $\alpha_1$.
Based on the obtained $\bz_{m}$, we determine the set of active modes $\mathcal{R}_{m}$ at time step $m$ as
\begin{align}
    \mathcal{R}_{m} = \left\{ j \in \{1,...,r\} \,\middle|\, \lambda_{j} \in \mathbb{C} \setminus \{0\},\ |z_{jm}| > 0 \right\}. \label{Eq:Mode Set}
\end{align}
This set collects the indices $\{ j \}$ of the non-zero eigenvalues whose corresponding $\{ z_{jm} \}$ are non-zero at time step $m$ (we introduce a small tolerance value $10^{-9}$ in the actual numerical analysis).
It captures the active modes that contribute to the dynamics at each time step, while ignoring the inactive modes.
The cardinality of $\mathcal{R}_{m}$, i.e., the number of active modes at time step $m$, reflects local complexity of the dynamics, and the evolution of $\mathcal{R}_{m}$ over time helps us to detect transitions or transient phenomena.

Here, we clarify the meaning of a mode that dynamically contributes to the system's behavior.
In the present framework, a mode is regarded as dynamically contributing at time step $m$ if it is selected as an active component in $\mathcal{R}_m$ through the regularized reconstruction problem. 
Thus, the contribution is defined with respect to the assumed representation in Eq.~\eqref{Eq:Proposed Model}, in which the data are reconstructed by a sparse superposition of DMD modes and slowly varying amplitudes with fixed frequencies and time-varying amplitudes.

Next, we determine the values of $z_{jm}$ for all $j\in\mathcal{R}_{m}$ under the sparsity structure determined in the first step by solving the following constrained convex optimization problem without the $\ell_1$-penalty. 
For the case without projection, Eq.~\eqref{Eq:Opt Problem For z}, we solve
\begin{align}
    \begin{split}
    \bz_{m} & = \underset{\by\in\mathbb{C}^{r}}{\mathrm{argmin}}\left\|\bx_{m} - \bPhi
    \mathrm{diag}(\by)
    \begin{bmatrix}
        e^{i m \omega_{1}} \\ \vdots \\ e^{i m \omega_{r}}                               
    \end{bmatrix}\right\|_{2}^{2}\\
    & +  \alpha_{2}\left\|\by-\bz_{m-1}\right\|_{2}^{2} \\
    & \ \ \ \ \ \mathrm{s.t.}\ \bR_{m}^{\top}\by = \bm{0},\ \ m=\{0,1,...,M\},
\end{split}
\label{Eq:Opt Problem For z modified}
\end{align}
where the matrix $\bR_{m}\in\mathbb{R}^{r\times (r-|\mathcal{R}_{m}|)}$ in the constraint is determined from the set $\mathcal{R}_{m}$.
The columns of $\bR_{m}$ are the standard unit vectors in $\mathbb{R}^{r}$ where the position of $1$ corresponds to the components of the non-active modes in $\mathcal{R}_{m}$~\cite{spDMD}.
For example, when $r=5$ and $\mathcal{R}_{20}=\{1,3,5\}$ at $m=20$, the matrix $\bR_{20}$ is given as
\begin{align*}
    \bR_{20} = 
    \begin{bmatrix}
        0 & 0 \\
        1 & 0 \\
        0 & 0 \\
        0 & 1 \\
        0 & 0 \\
    \end{bmatrix}.
\end{align*}
This constraint imposes a sparsity structure on the solution of Eq.~\eqref{Eq:Opt Problem For z proj} given
by $\mathcal{R}_{m}$.

By using the above two-step method, we solve our optimization problem with the sparsity and smoothness regularization.
Note that the two-step procedure is necessary because $\ell_1$ regularization in the first step introduces a well-known amplitude bias~\cite{PostLASSO}; the penalized solution systematically underestimates the true amplitudes. By re-solving the problem in the second step with $\ell_1$ penalty removed and inactive modes constrained to zero (via $\bR_m^\top \by = \bm{0}$), we recover physically accurate, unbiased amplitude values for the selected active modes.
The constraint $\bR_m^\top \by = \bm{0}$ is enforced by restricting the optimization variables to the $|\mathcal{R}_m|$ active components. That is, the inactive components are fixed at zero, and the optimization is performed only over the active components, which reduces the problem to a standard unconstrained least-squares minimization for the active modes. 
In Sec.~\ref{Sec:SimpleExample}, we show that this two-step procedure is crucial for accurately capturing the transient dynamics of the system.
We note that the active DMD modes are chosen at each time step to satisfy the sparsity and smoothness, in contrast to the sparsity-promoting DMD that chooses the active DMD modes over the whole time domain~\cite{spDMD}.

\section{Frequency perspective}\label{Sec:FrequencyPerspective}
%%%%%%%%%%%%%%%%%%%%%%%%%%%%%%%%%%%%%%%%%%%%%%%%%%%%%%%%
%%% Frequency perspective %%%%%%%%%%%%%%%%%%%%%%%%%%%%%%
%%%%%%%%%%%%%%%%%%%%%%%%%%%%%%%%%%%%%%%%%%%%%%%%%%%%%%%%
In this section, we discuss the method from a frequency domain perspective, which provides a clearer understanding of the spectral properties of the coefficients of the extracted modes.

\subsection{Physical interpretation of the modes}

We begin by showing that the time evolution of each DMD mode can be interpreted as an amplitude-modulated and frequency-modulated (AM--FM) signal, analogous to the Intrinsic Mode Functions (IMFs)~\cite{EMD2,EWT,VMD}. 
The IMFs provide a useful framework for decomposing complex, non-stationary, and nonlinear signals. 
Each IMF represents a well-defined and physically interpretable oscillatory component with time-varying amplitude and frequency~\cite{VMD,EMD1}.
It can be written in the form~\cite{EMD2}
\begin{align}
    u(t) = F(t) \cos \varphi(t),
    \ 
    \quad  
    F(t), \dot{\varphi}(t) \geq 0. 
\end{align}
The main assumption is that $F$ and $\dot{\varphi}$ vary much slower than $\varphi$.
The IMF $u$ behaves as an amplitude-modulated harmonic signal.

The decomposition in Eq.~\eqref{Eq:Proposed Model} may be written in continuous time as
\begin{align}
     \bx(t) \approx \sum_{j=1}^{r}e^{i\Omega_{j}t}z_{j}(t)\bphi_{j}, \label{Eq:Proposed Model continuous}
\end{align}
where $\Omega_{j}$ is a continuous-time frequency, i.e., $\Omega_{j}=\omega_{j}/\Delta t$
and $\mathcal{R}(t)$ is the set of active modes at time $t$.
We denote the coefficient $s_{j}(t)$ of the $j$th DMD mode as 
\begin{align}
    s_{j}(t) = z_{j}(t)e^{i\Omega_{j}t}, \label{Eq:Signal}
\end{align}
which represents an amplitude-modulated harmonic oscillation of frequency $\Omega_{j}$ with the amplitude $z_j(t)$.
By expressing the complex amplitude in the polar form, $z_{j}(t)=|z_{j}(t)|e^{i\theta_{j}(t)}$, the real part of $s_j(t)$ can be written as
\begin{align*}
    \mathfrak{R}[s_{j}(t)] = |z_{j}(t)|\cos\left(\Omega_{j}t + \theta_{j}(t)\right).
\end{align*}
This expression has the form of an AM--FM signal similar to IMF, where $|z_{j}(t)|$ is the instantaneous amplitude and $\Omega_{j}+\dot{\theta}_{j}(t)$ is the instantaneous frequency.  

For this analogy to IMF to be meaningful, the signal should satisfy the following conditions~\cite{EMD2,EWT,VMD}:
\begin{itemize}
    \item The phase $\Omega_{j}t + \theta_{j}(t)$ is a non-decreasing function, $\Omega_{j}+\dot{\theta}_{j}(t)\geq0$.
    \item Both the amplitude $|z_{j}(t)|$ and the instantaneous frequency $\Omega_{j}+\dot{\theta}_{j}(t)$ vary much slower than the phase $\Omega_{j}t + \theta_{j}(t)$.
\end{itemize}
If these conditions are satisfied due to the smoothness constraint on $z_j(t)$, locally, i.e., on a time interval $[t_{0}-\delta, t_{0}+\delta]$, with $\delta\approx2\pi[\Omega_{j}+\dot{\theta}_{j}(t_{0})]^{-1}$, the signal $\mathfrak{R}[s_{j}(t)]$ can be regarded as a pure harmonic signal with amplitude $|z_{j}(t_{0})|$ and instantaneous frequency $\Omega_{j}+\dot{\theta}_{j}(t_{0})$.
In particular, when the rate of change of the frequency, $\dot{\theta_{j}}(t_{0})$, is sufficiently small compared to the base frequency $\Omega_{j}$, i.e., $\dot{\theta}_{j}\ll\Omega_{j}$, we can regard that the $j$th DMD mode of the system exhibits harmonic oscillations with a nearly constant amplitude $|z_{j}(t_{0})|$ and frequency $\Omega_{j}$ in this local interval.
Thus, each DMD coefficient can be locally interpreted as a simple pure tone with a clear physical meaning. 

\subsection{Smoothness and spectral spread}

We now explain that the smoothness constraint adopted in our method promotes spectral compactness of the coefficients of the active modes.
Equation~\eqref{Eq:Opt Problem For z modified} incorporates a smoothness constraint on the amplitude to capture the underlying dynamics without introducing spurious, abrupt changes of the amplitudes. 
This temporal smoothness has a direct consequence in the frequency domain, which leads to the IMF-like properties of the extracted modes discussed above.

We first quantify the non-smoothness of the amplitude $z_j(t)$ using the squared integral of its time derivative as
\begin{align*}
D_{j} = \int \left|\frac{dz_{j}(t)}{dt}\right|^2 dt.
\end{align*}
Note that a discrete version of this quantity is included in our optimization problem as the regularization term for smoothness.
Using Parseval's theorem, $D_j$ can be expressed in the frequency domain as
\begin{align}
D_{j} = \frac{1}{2\pi}\int_{-\infty}^{\infty}\Omega^2\left|\hat{z}_{j}(\Omega)\right|^2d\Omega, \label{Eq:Continuous_Dj_Freq}
\end{align}
where $\hat{z}_j(\Omega)$ is the Fourier transform of $z_j(t)$.
This equation reveals that minimizing $D_j$ in the time domain is equivalent to penalizing the high-frequency components of the amplitude, thus enforcing the slowly varying property.

Next, we evaluate the spectral spread of $s_j(t) = z_j(t)e^{i\Omega_j t}$ around the base frequency $\Omega_j$ using the spectral variance as
\begin{align*}
\sigma^{2}_{j} = \frac{\displaystyle\int_{-\infty}^{\infty}(\Omega-\Omega_{j})^2\left|\hat{s}_{j}(\Omega)\right|^2d\Omega}{\displaystyle\int_{-\infty}^{\infty}\left|\hat{s}_{j}(\Omega)\right|^2d\Omega}.
\end{align*}
Here, $\hat{s}_j(\Omega)$ is the Fourier transform of $s_j(t)$ satisfying
\begin{align*}
    \hat{s}_{j}(\Omega) = \hat{z}_{j}(\Omega)*\delta(\Omega - \Omega_j) = \hat{z}_{j}({\Omega - \Omega_{j}}),
\end{align*}
where $\delta(\cdot)$ is the Dirac delta function and $*$ denotes convolution. 
Thus, we obtain
\begin{align}
\sigma^{2}_{j} = \frac{\displaystyle\int_{-\infty}^{\infty}\Omega'^2\left|\hat{z}_{j}(\Omega')\right|^2d\Omega'}{\displaystyle\int_{-\infty}^{\infty}\left|\hat{z}_{j}(\Omega')\right|^2d\Omega'}.
\label{Eq:Continuous_Variance}
\end{align}

By comparing Eqs.~\eqref{Eq:Continuous_Dj_Freq} and \eqref{Eq:Continuous_Variance}, we establish a direct relationship between the temporal smoothness of the amplitude and the spectral spread of the mode,
\begin{align}
\sigma_j^2 = \frac{D_j}{E_j}, \quad E_j = \int \left|z_j(t)\right|^2 dt,
\end{align}
where the denominator $E_j$ is the total energy of the amplitude. 
This result shows that minimizing the variation $D_j$ of the amplitude, which is the objective of the smoothness constraint in our method, directly minimizes the spectral variance $\sigma_j^2$.

Thus, the smoothness constraint in our proposed method is not just an ad hoc addition, but it is a mechanism that ensures that the extracted coefficients of the DMD modes are spectrally compact and behave like well-defined, physically interpretable IMFs.

\section{Simple Example}  \label{Sec:SimpleExample}

%%%%%%%%%%%%%%%%%%%%%%%%%%%%%%%%%%%%%%%%%%%%%%%%%%%%%%%%
%%% Simple Example %%%%%%%%%%%%%%%%%%%%%%%%%%%%%%%%%%%%%
%%%%%%%%%%%%%%%%%%%%%%%%%%%%%%%%%%%%%%%%%%%%%%%%%%%%%%%%
\begin{figure}[h]
    \centering
    \includegraphics[scale=.35]{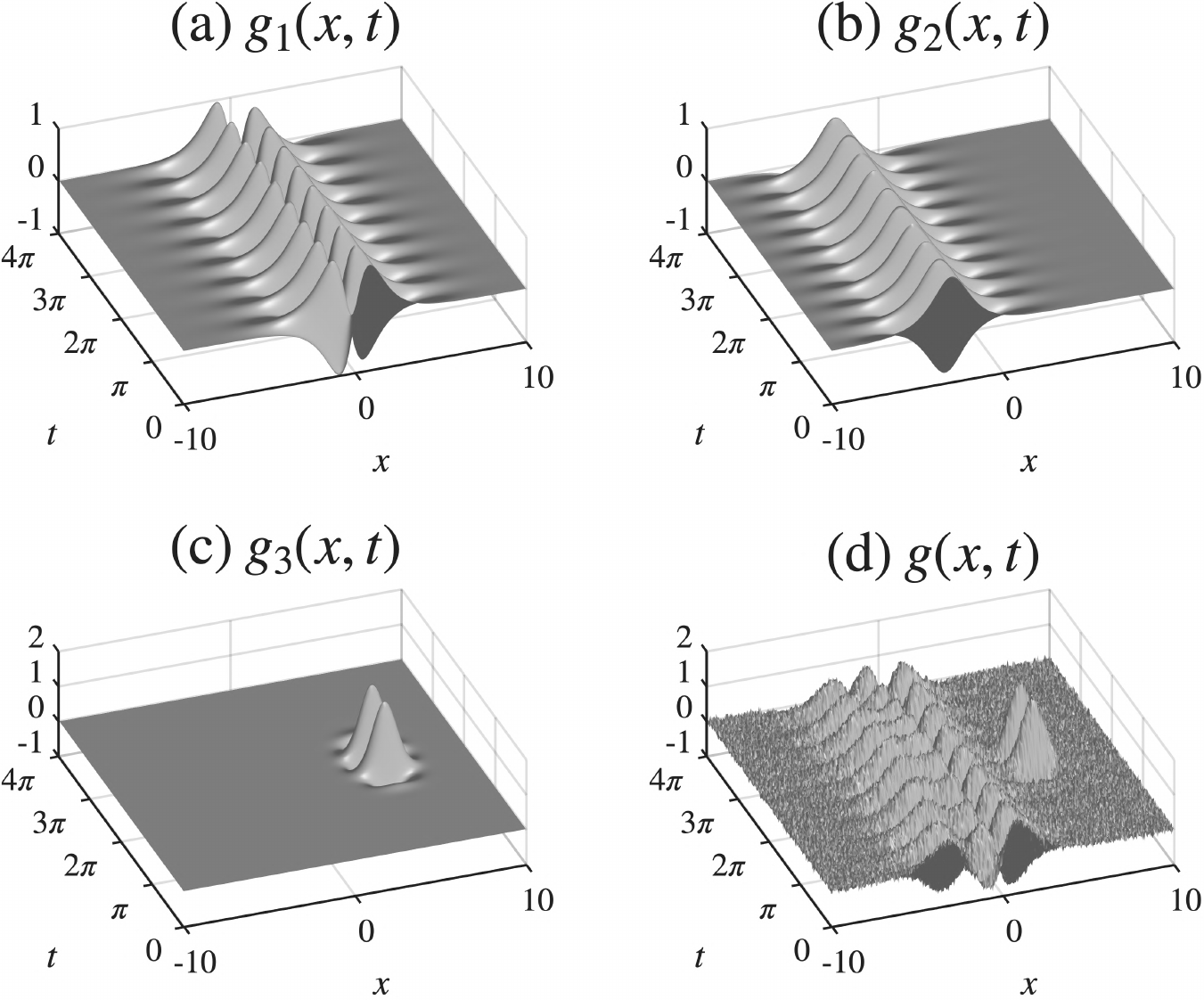}
    \caption{
    Example of three mixed spatiotemporal signals.
    (a) $g_1(x,t)$, (b) $g_2(x,t)$, (c) $g_3(x,t)$, and (d) $g(x,t) = g_1+g_2+g_3 +n$ defined in Eq.~\eqref{Eq:Spatiotemporal Signals}.
    Only the real parts are plotted.
    }
    \label{Fig:Simple Example 1}
\end{figure}

\begin{figure*}[t]
    \centering
    \includegraphics[scale=.58]{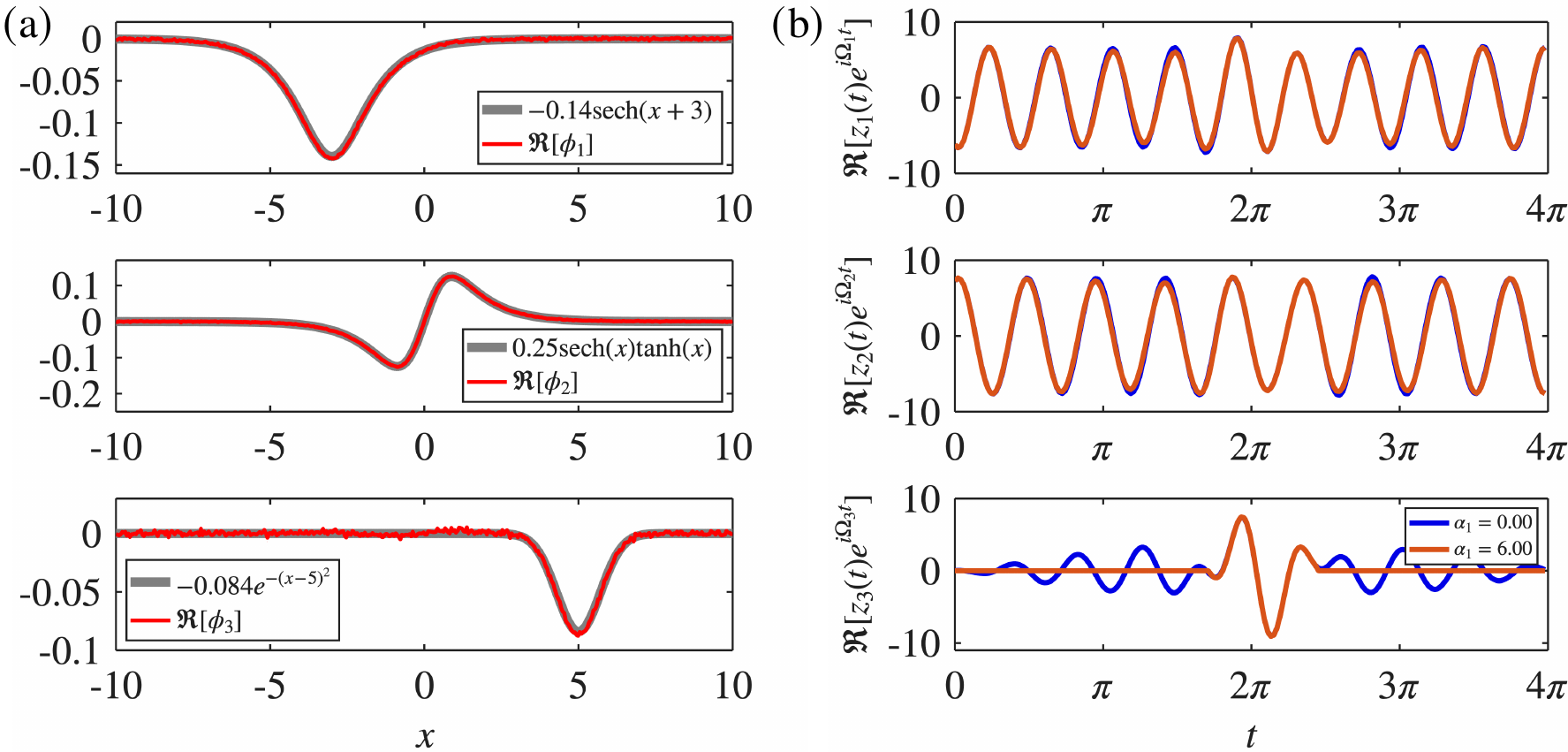}
    \caption{{\bf
    Effect of sparsity regularization.
    Results of the proposed method for the spatiotemporal signal in Fig.~\ref{Fig:Simple Example 1}. (a)~Red dashed lines represent the real part of the DMD modes $\{ \bphi_{j} \}$. 
    Gray lines represent the spatial profiles of the three components $g_{1}$, $g_{2}$, and $g_{3}$, respectively (rescaled to fit the DMD modes).   
    (b)~Real part of the extracted coefficient $z_{j}(t)e^{i\Omega_{j}t}$ of the DMD mode $j$ ($j=1,2,3$). The blue and orange lines correspond to $\alpha_{1}=0$ and $\alpha_{1}=6$, respectively. $\alpha_{2}=4$ in both cases.
    Graphs showing the results of the proposed method applied to the spatiotemporal signal. (a) A comparison between the real part of the extracted DMD modes (red dashed lines) and the spatial profiles of the original signal components (gray lines). (b) The temporal evolution of the real part of the coefficients for each DMD mode, comparing cases without sparsity regularization (blue line) and with sparsity regularization (orange line).
    }}
    \label{Fig:Simple Example 2}
\end{figure*}

To illustrate the proposed method, we first apply it to a simple example of three mixed spatiotemporal signals taken from Ref.~\cite{DMD Book}. 
Using this example, we demonstrate three key aspects: (i)~the effect of sparsity regularization in identifying dynamically active modes, (ii)~the role of smoothness  regularization in controlling the spectral compactness of the amplitudes, and (iii)~the necessity of the two-step procedure for obtaining unbiased amplitude estimates.
In this example, the spatiotemporal signal is synthesized as
\begin{align}
    \begin{split} \label{Eq:Spatiotemporal Signals}
    g(x,t) =& g_1(x,t) + g_2(x,t) + g_3(x,t) + n(x,t)\\
    = & \mathrm{sech}(x + 3) \exp(i4.3t) \\
    &+ 2 \mathrm{sech}(x) \tanh(x) \exp(i4.8t) \\
    &+ 3\exp[-\{(x - 5)^2 + (t - \pi )^2\}]\exp(i4.5t)\\
    &+n(x,t),
    \end{split}
\end{align}
for $x\in[-10,10]$ and $t\in[0,4\pi]$, where $n(x,t)$ denotes the added complex white Gaussian noise with zero mean and variance $\sigma^2$.
Its real and imaginary parts independently follow a normal distribution such that $\Re[n],\Im[n]\sim\mathcal{N}(0,\sigma^{2}/2)$ with $\sigma=0.1$.
The two signals $g_1$ and $g_2$ are plotted in Figs.~\ref{Fig:Simple Example 1}~(a) and (b).
They represent spatially localized oscillatory components with constant amplitudes and continuous-time frequencies $\Omega_{1}=4.3$ and $\Omega_{2}=4.8$, respectively. 
The signal $g_3$ is plotted in Fig.~\ref{Fig:Simple Example 1}~(c), which represents a transient component that is localized in both space and time around $(x,t)=(5, 2\pi)$, with a continuous-time frequency of $\Omega_{3}=4.5$.
The mixed signal $g=g_{1}+g_{2}+g_{3} + n$ is illustrated in Fig.~\ref{Fig:Simple Example 1}~(d).

To obtain the snapshots of the mixed signals $g$, we discretize the spatial domain $x\in[-10,10]$ into $N=400$ grid points and the time domain $t\in[0,4\pi]$ into $M=200$ grid points. Then, we arrange the obtained snapshots into the data matrices $\bX_{-}$ and $\bX_{+}$ as in Eq.~\eqref{Eq:Data Matrix} and apply the rank-3 exact DMD to extract the DMD modes $\{ \bphi_{j} \}$ and the continuous-time eigenvalues $\{\mu_j\} = \{\log(\lambda_j)/\Delta t\}$.
The noise is known to introduce systematic bias in the estimated eigenvalues, which leads to underestimation of the magnitude of the eigenvalues and overestimation of the decay rates~\cite{optDMD,noiseDMD}.
The estimated eigenvalues are $\mu_1 = -0.0043 + 4.2974i$, $\mu_2 = -0.0027 + 4.8013i$, and $\mu_3 =-0.0623 + 4.4602i$, whose imaginary parts closely match the true frequencies $\{\Omega_j\}$.
Figure~\ref{Fig:Simple Example 2}~(a) shows that the real parts of the extracted DMD modes accurately reproduce the spatial profiles of the signals $g_1$, $g_2$, and $g_3$.

Thus, this simple example demonstrates that the proposed sparsity regularization can adaptively identify when each mode is dynamically active, which is impossible with the standard DMD.
In what follows, we discuss the three aspects of the proposed method, i.e., the sparsity, smoothness, and two-step procedure.

\paragraph*{(i)~Effect of sparsity regularization (with $\alpha_{2}$ fixed).}
We first examine the role of the sparsity parameter $\alpha_{1}$ with $\alpha_{2}=4$ fixed.
Figure~\ref{Fig:Simple Example 2}~(b) shows the real part of $z_{j}(t)e^{i\Omega_{j}t}$ for $j=1,2,3$.
The blue line ($\alpha_1=0$) corresponds to the case without sparsity regularization.
In this case, all three modes attain non-zero amplitudes at every time step, so the transient nature of $g_3$ is not captured clearly. 
The third-mode coefficient oscillates throughout the entire time horizon rather than being localized near $t = 2\pi$.
In contrast, the orange line ($\alpha_1=6$) with sparsity regularization successfully recovers the underlying structure. 
The first two modes remain active at all times, while the amplitude of the third mode is non-zero only near $t = 2\pi$ and vanishes elsewhere, accurately reflecting the true localized pulse $g_{3}$ in Fig.~\ref{Fig:Simple Example 1}.

\begin{figure}[htbp]
    \centering
    \includegraphics[scale=.42]{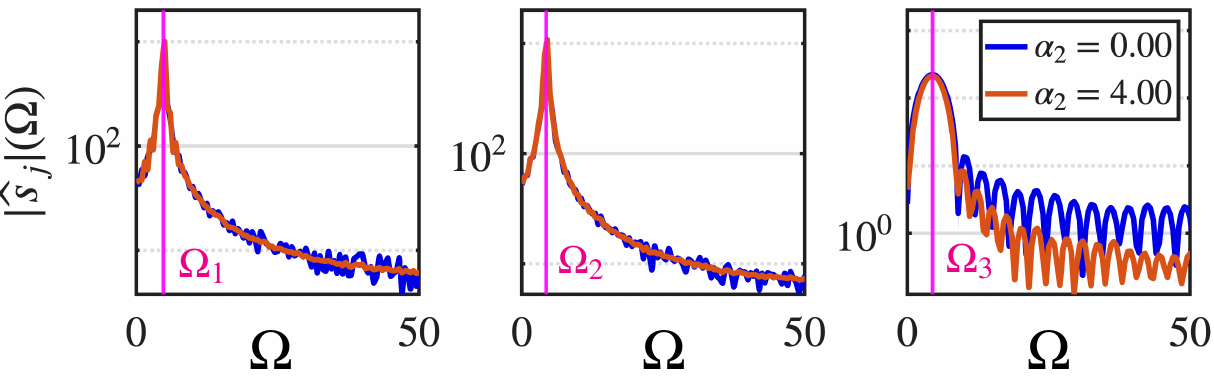}
    \caption{
    Effect of smoothness regularization. 
    Power spectra $|\hat{s}_j|(\Omega)$ of the extracted coefficients $s_j(t)=z_j(t)e^{i\Omega_j t}$ for the three DMD modes.
    The blue and orange lines correspond to $\alpha_2=0$ and $\alpha_2=4$, respectively, with $\alpha_1=6$ fixed.
    The magenta vertical lines indicate the DMD eigenfrequencies $\Omega_j$.}
    \label{Fig:Simple Example 3}
\end{figure}

\paragraph*{(ii)~Effect of smoothness regularization (with $\alpha_{1}$ fixed).}We next examine the role of the smoothnessonly  parameter $\alpha_{2}$ with $\alpha_{1}=6$ fixed.
Figure~\ref{Fig:Simple Example 3} shows the power spectra $|\hat{s}_j|(\Omega)$
of the extracted coefficients $s_j(t) = z_j(t)e^{i\Omega_j t}$ for $j = 1, 2, 3$, plotted against continuous-time frequency $\Omega$.
The blue and orange lines correspond to $\alpha_2 = 0$ and $\alpha_2 = 4$, respectively, and the vertical magenta lines indicate the eigenfrequency $\Omega_j$ of each mode.

For the first and second modes, the effect of increasing $\alpha_2$ is limited.
This is because the true components $g_{1}$ and $g_{2}$ have constant amplitudes and remain active throughout the whole time interval.
As a result, even without the smoothness regularization, the estimated coefficients are already slowly varying amplitude-modulated signals and their spectra are concentrated around the corresponding eigenfrequencies.

In contrast, the effect of $\alpha_{2}$ is more clearly observed for the third mode.
The true component $g_3$ is localized in time and space, so its coefficient is more sensitive to observation noise and to small errors in the DMD modes and eigenfrequencies.
When $\alpha_{2}=0$, these effects appear as high-frequency fluctuations in the estimated coefficient, which leads to a broad spectrum of $s_{3}(t)$ around $\Omega_{3}$.
By imposing the smoothness regularization, these spurious high-frequency fluctuations are suppressed, and the spectrum becomes more concentrated around the eigenfrequency.

\begin{figure}[h]
    \centering
    \includegraphics[scale=.43]{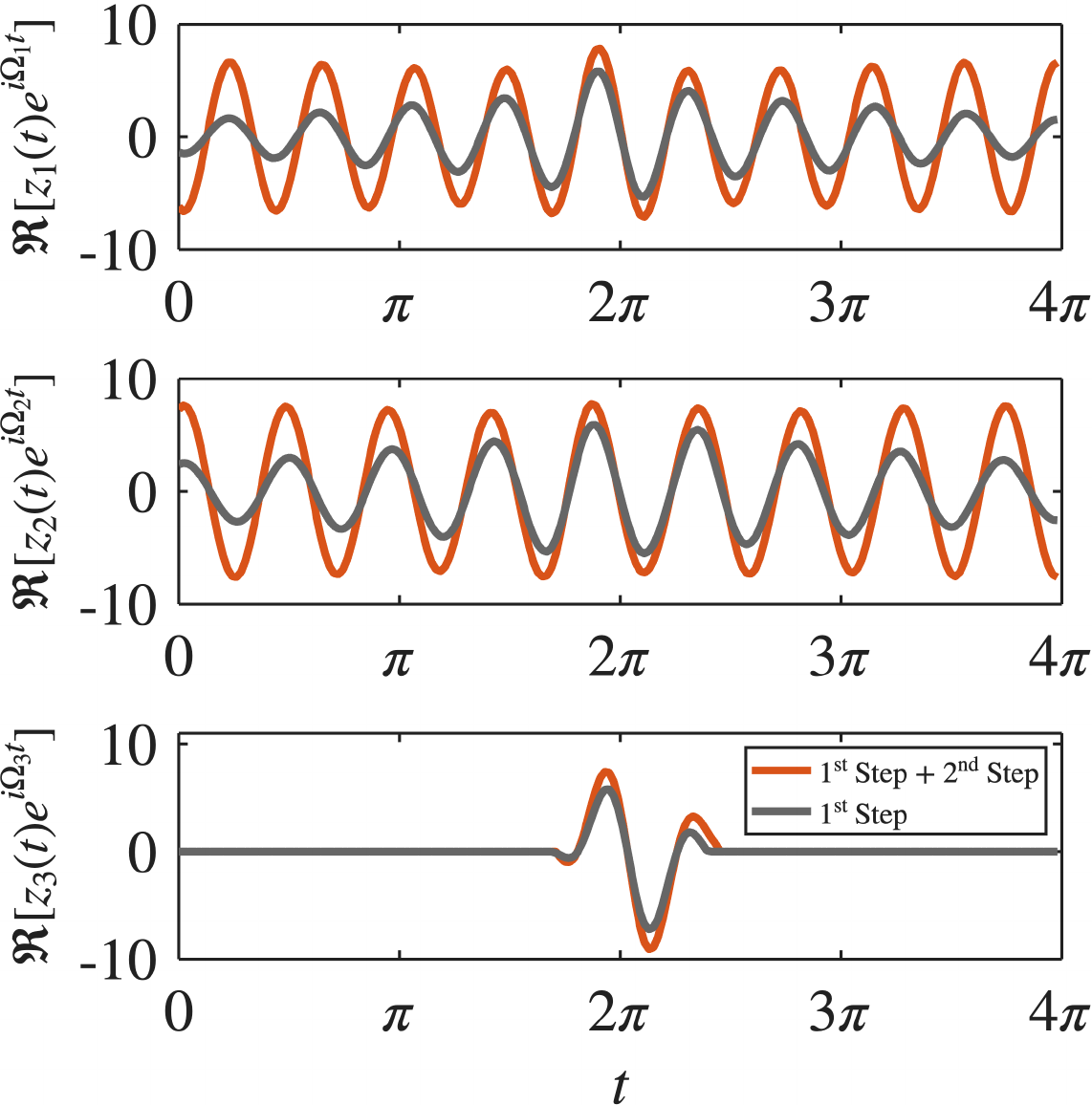}
    \caption{
    Importance of the two-step procedure. 
    Comparison of the coefficients obtained after only the first $\ell_1$ regularized step (gray lines) and after the full two-step procedure (orange lines).
    Graphs illustrating the importance of the two-step procedure by comparing the extracted mode coefficients. The gray lines represent the coefficients obtained after only the first L1 regularized step, while the orange lines represent the recovered coefficients after the full two-step procedure.}
    \label{Fig:Simple Example 4}
\end{figure}

\paragraph*{(iii)~Importance of the two-step procedure.}
Finally, we illustrate the necessity of the two-step procedure described in Sec.~\ref{Sec:Method}.
Figure~\ref{Fig:Simple Example 4} compares the coefficients obtained after only the first step (gray lines) with those obtained after both the first and second steps (orange lines).
In the first step, the sparsity structure is determined by solving the $\ell_1$ regularized problem in Eq.~\eqref{Eq:Opt Problem For z}, which is effective in identifying the active set $\mathcal{R}_{m}$ but introduces an amplitude bias due to the $\ell_{1}$ penalty.
This amplitude bias is observed in Fig.~\ref{Fig:Simple Example 4}. 
When only the first step is used, the extracted coefficients underestimate the amplitudes of the true components.
Therefore, the first step can select the active modes, but it does not recover the true amplitudes accurately. 

In contrast, the second step removes the $\ell_1$ penalty and solves a constrained optimization problem that enforces the sparsity structure determined in the first step.
As shown in Fig.~\ref{Fig:Simple Example 4}, this procedure successfully recovers the amplitudes more accurately with the same sparsity structure preserved.
Thus, the two-step procedure is essential for the unbiased amplitude estimation of the active modes while maintaining the sparsity structure that captures the transient dynamics.

\section{Numerical Experiment} \label{Sec:NumericalExperiment}

%%%%%%%%%%%%%%%%%%%%%%%%%%%%%%%%%%%%%%%%%%%%%%%%%%%%%%%%
%%% Numerical Experiment %%%%%%%%%%%%%%%%%%%%%%%%%%%%%%%
%%%%%%%%%%%%%%%%%%%%%%%%%%%%%%%%%%%%%%%%%%%%%%%%%%%%%%%%
In this section, we apply the proposed method to a dataset of numerically simulated fluid flow to analyze the transient dynamics. 

\begin{figure}[t]
    \centering
    \includegraphics[scale=.36]{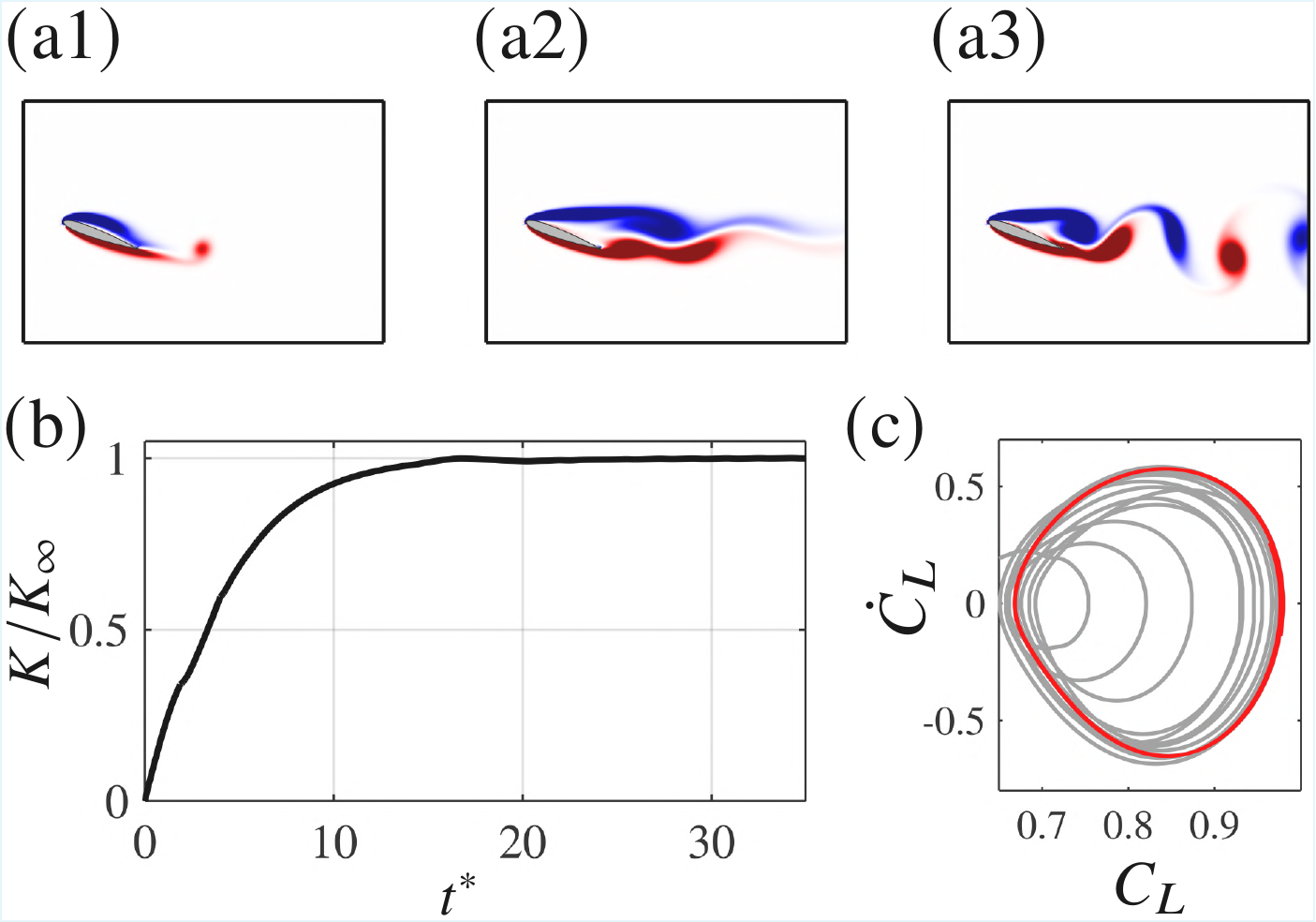}
    \caption{
    K\'arm\'an vortex street.
    (a) Three snapshots of the vorticity field at the initial ($t^{*}=2$), intermediate ($t^{*}=9$), and late ($t^{*}=30$) stages. 
    (b) Time evolution of the normalized kinetic energy $K/K_{\infty}$, where $K$ is the total kinetic energy at time $t^{*}$ and $K_{\infty}$ is the time-averaged kinetic energy in the steady limit-cycle state.
    (c) Lissajous plot ($C_L$ vs $\dot{C}_L$) showing the transient trajectory (gray) converging to the steady limit cycle (red).
    Simulation results of the fluid flow. (a) Visualizations of the vorticity field at the initial, intermediate, and late stages. (b) A graph showing the time evolution of the normalized kinetic energy. (c) A Lissajous plot of the lift coefficient versus its time derivative, showing the transient trajectory (gray line) converging to the steady limit cycle (red line).
    }
    \label{Fig:Simulation Result}
\end{figure}

\subsection{Setup}

As the dataset, we have chosen a two-dimensional incompressible viscous flow through an NACA0012 airfoil with an angle of attack $20^{\circ}$, where the Reynolds number based on the wing chord is $500$. 
We denote the chord length as $c$ and set the full computational domain to be $(x,y)/c\in[-2,12]\times[-6,6]$.
The region of interest from which vorticity snapshots are extracted is $(x,y)/c\in[-0.5,4]\times[-1.5,1.5]$.
The leading edge of the airfoil is positioned at the origin.
We simulated the flow using the lattice-Boltzmann method with the D2Q9 model~\cite{LBM1,LBM2}.
The airfoil boundary is treated with the standard bounce-back condition (no-slip), and vorticity values inside the airfoil are set to zero (zero-masking) in the post-processing.

An oscillating pattern of the vortices, i.e., K\'arm\'an vortex street, is formed by the airfoil placed against the fluid flow, which can be considered a limit-cycle solution of the system. 
We confirmed that flows starting from a certain range of initial conditions approach this limit-cycle as time progresses.
Three typical vorticity snapshots are shown in Fig.~\ref{Fig:Simulation Result}(a), where (a1), (a2), and (a3) correspond to the initial, intermediate, and late stages of the flow, respectively. 
In the initial stage, a coherent wake behind the airfoil begins to develop, yet no clear vortex shedding is observed. 
During the intermediate stage, the wake gradually becomes asymmetric, leading to the onset of alternating vortex shedding. 
Finally, a stable K\'arm\'an vortex street is established, indicating that the flow converges to the steady limit-cycle dynamics.

To demonstrate the convergence, we calculated the total kinetic energy $K$ of the flow field at each time step normalized by the time-averaged kinetic energy $K_{\infty}$ in the steady limit-cycle state. 
The time evolution of the normalized kinetic energy is shown in Fig.~\ref{Fig:Simulation Result}(b), demonstrating the convergence toward the steady limit-cycle state. 
Moreover, Fig.~\ref{Fig:Simulation Result}~(c) shows the Lissajous plot of the lift coefficient $C_{L}$ vs its time derivative $\dot{C}_{L}$, where the gray and red lines represent the transient trajectory and the steady limit cycle, respectively.

All results are reported in convective time units $t^{*} = t\,u_{\infty}/c$, where each snapshot index $m$ corresponds to $t^{*}=m\Delta t$.
The simulation
provides $M=700$ vorticity snapshots $\{\bm{x}_{m}\}_{m=0}^{700}$ sampled at  a fixed time interval $\Delta t = 0.05\, c/u_{\infty}$, covering the full trajectory from $t=0$ to $t=35\, c/u_{\infty}$. 
Each snapshot has $284 \mbox{(vertical)}\times 425 \mbox{(horizontal)}$ measurement points in the region of interest.
The Courant--Friedrichs--Lewy (CFL) number in the simulation is $0.08$~\cite{LBM1,LBM2}.

\begin{figure*}[t]
    \centering
    \includegraphics[scale=.76]{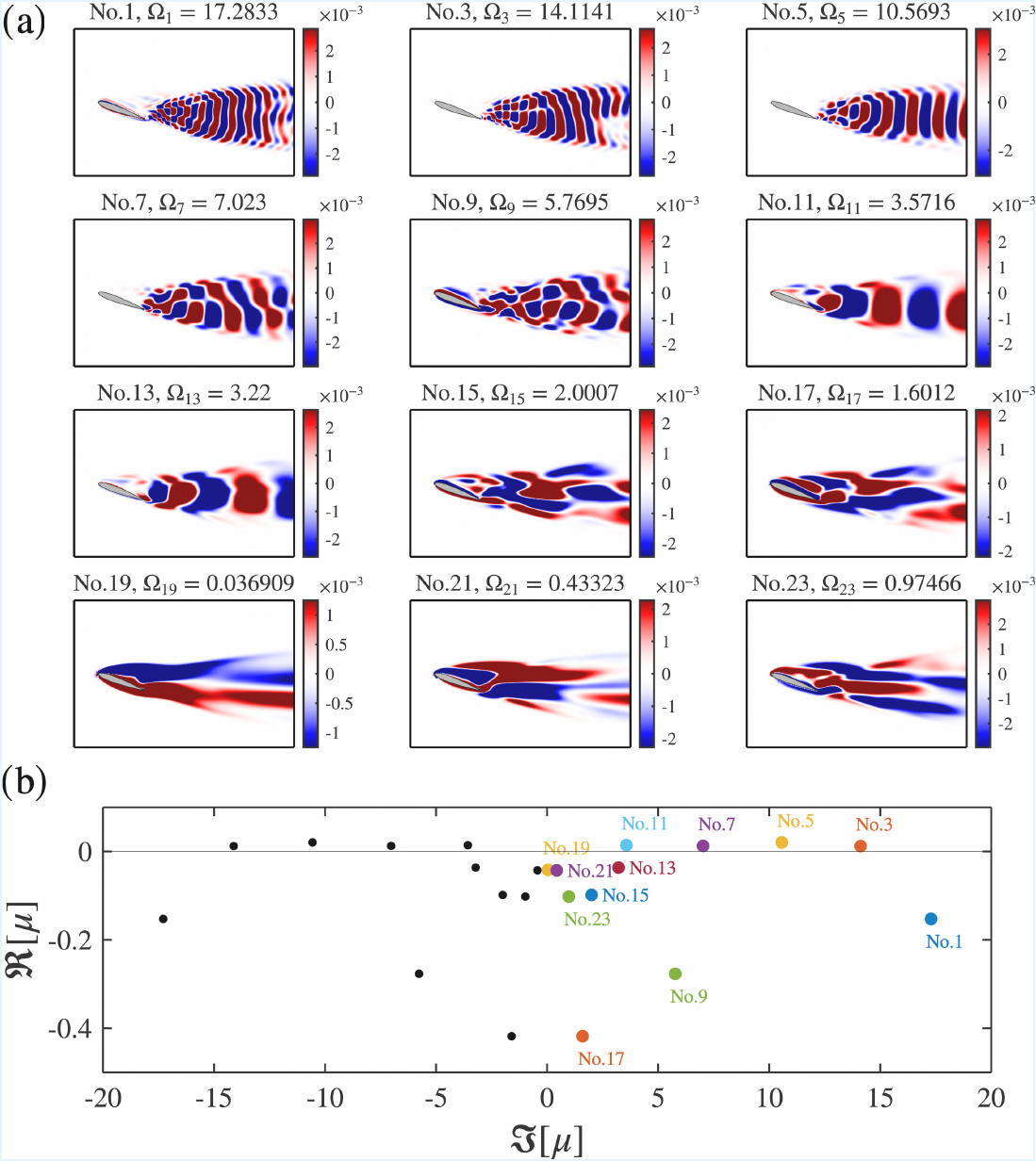}
    \caption{{%\bf
    Results of exact DMD with $r=24$.
    (a) Vorticity of the $j$th DMD mode ($j=1,3,5,7,9,11,13,15,17,19,21$, and $23$) visualized in color code, with the continuous-time frequency $\Omega_j = \omega_j/\Delta t$ shown for each mode.
    Only the odd modes are shown, since the even modes are complex conjugates of the odd modes.
    (b) Plot of the continuous-time DMD eigenvalues $\{\mu_j\}$ on the complex plane, where $\mu_j = \log(\lambda_j)/\Delta t$ and $\lambda_j$ is the discrete-time eigenvalue. The dashed line indicates $\mathfrak{R}[\mu_j]=0$ (neutrally stable modes).
    Results of the Exact DMD application. (a) Images visualizing the vorticity of the extracted DMD modes using a color code. (b) A plot of the continuous-time DMD eigenvalues on the complex plane, with a dashed line indicating neutrally stable modes.
    }}
    \label{Fig:Exact DMD Result}
\end{figure*}

The $M=700$ snapshots were chosen to cover the complete dynamical process from the initial quiescent state, through the nonlinear transient, to the established limit-cycle regime, as confirmed by convergence of $K/K_\infty$ in Fig.~\ref{Fig:Simulation Result}(b) and the closure of the Lissajous plot in Fig.~\ref{Fig:Simulation Result}(c).
In addition, the eigenvalues $\mu_{j}$ of the modes of interest lie close to $\Re[\mu_j]=0$ (Fig.~\ref{Fig:Exact DMD Result}(b)), confirming the chosen time horizon produces DMD eigenvalues $\lambda_{j}$ sufficiently close to the unit circle for the slow-amplitude assumption of the proposed method to hold.
If a shorter time horizon is used, the decomposition may miss the late-stage 
attractor; if a longer one is used, redundant limit-cycle data are included without a qualitative change in the extracted transient structure, potentially masking the transient features.
In general, the data length should be determined based on whether the dataset contains the transient phenomenon of interest and whether the DMD eigenvalues of the modes of interest are sufficiently close to the unit circle, which can be checked by physical observables and the DMD eigenvalues, respectively.

\begin{figure*}[htbp]
    \centering
    \includegraphics[scale=.65]{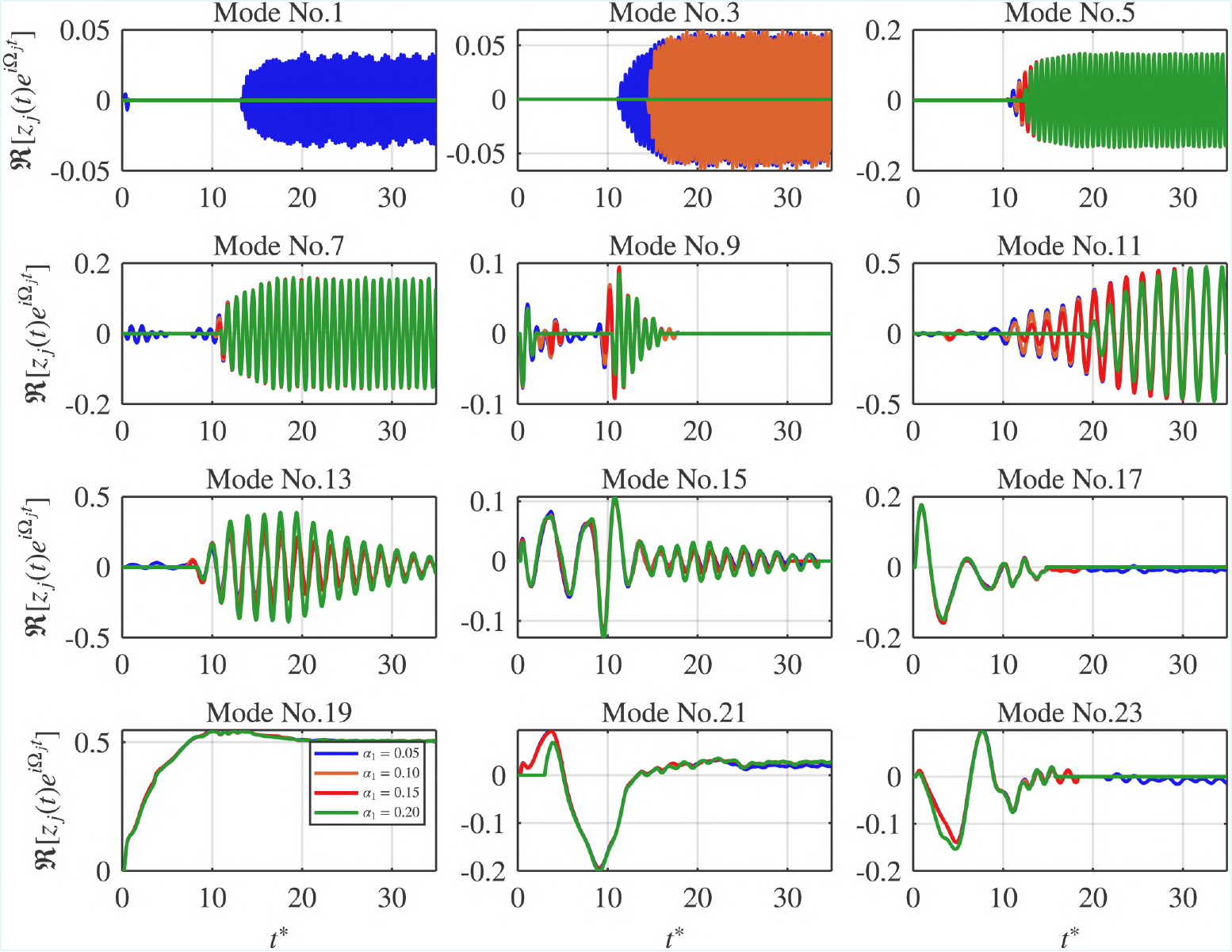}
    \caption{
    {%\bf
    Real part of the coefficient of each DMD mode vs convective time $t^{*}$.
    Lines represent $\mathfrak{R}[z_{jm}e^{i\omega_{j}m}]$ in Eq.~\eqref{Eq:Proposed Model} using $z_{jm}$ obtained from Eq.~\eqref{Eq:Opt Problem For z modified}, plotted against $t^{*}=m\Delta t$.
    The blue, orange, red, and green lines correspond to $\alpha_{1}=0.05$, $0.10$, $0.15$, and $0.20$, respectively, with all cases using $\alpha_{2} = 10$.
    A graph showing the real part of the coefficient of each DMD mode versus convective time. It compares the temporal evolution of the coefficients under multiple sparsity parameter settings represented by blue, orange, red, and green lines. }}
    \label{Fig:TemporalStructure}
\end{figure*}

\begin{figure*}[htbp]
    \centering
    \includegraphics[scale=.6]{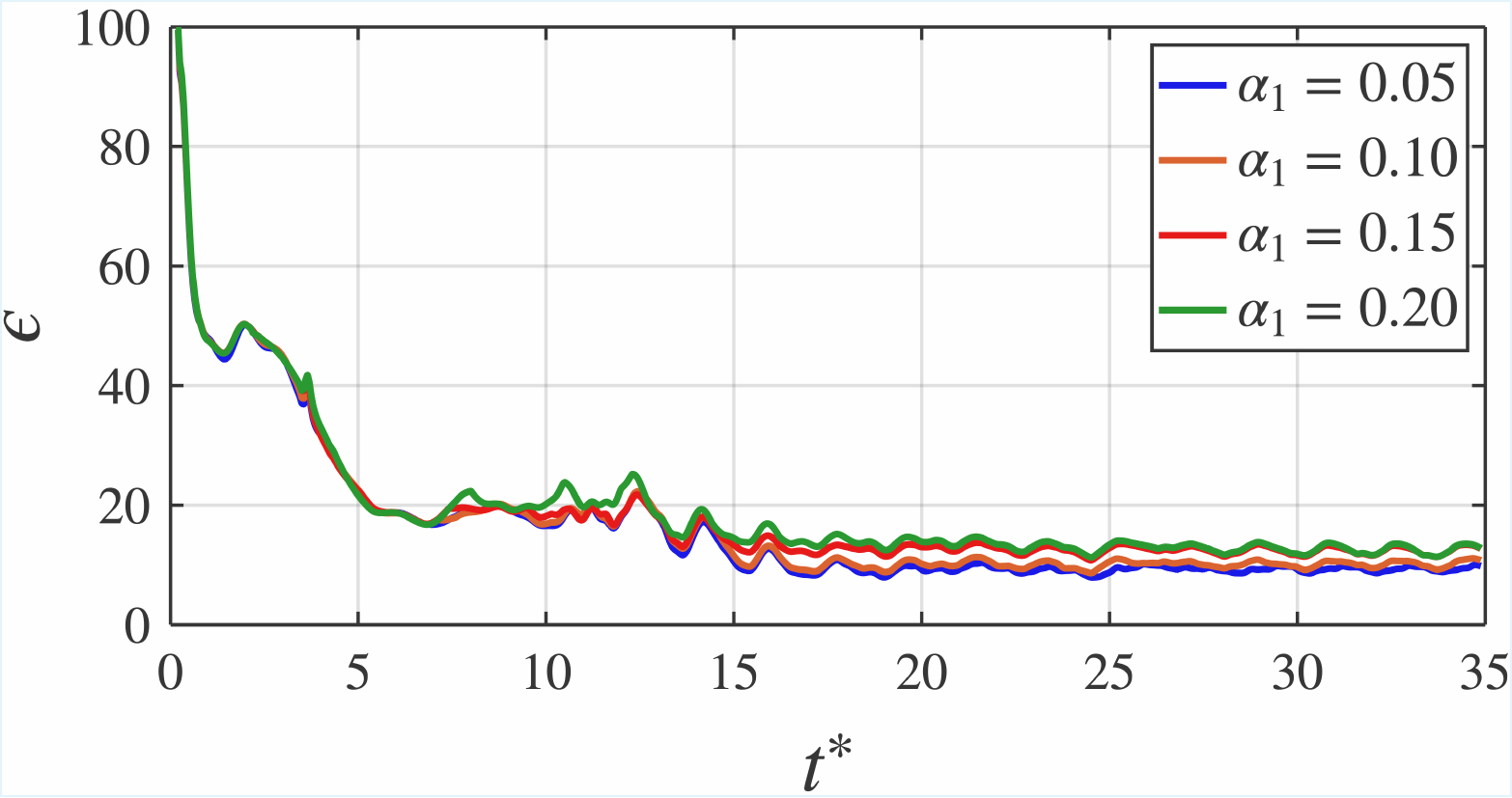}
    \caption{
    {%\bf
    Reconstruction error $\epsilon_m$ of the simulated data vs convective time $t^{*}$ for $\alpha_{1}=0.05$, $0.10$, $0.15$, and $0.20$ (all with $\alpha_2=10$).
    A graph showing the reconstruction error of the simulated data versus convective time. It illustrates how the error decays over time for different sparsity parameters.
    }}
    \label{Fig:ProjectionError}
\end{figure*}

\subsection{Results}

We apply the proposed method to extract transient activity of the DMD modes from time-series data. 
For fluid flows, previous studies have shown that DMD can accurately capture the steady behavior on the attractor after the initial transient~\cite{DMD,Koopman analysis flow1,Koopman analysis flow3,DDL}. 
However, several studies, including Noack {\em et al}.~\cite{RDMD} and Bagheri~\cite{KMD}, have 
also pointed out the difficulties in the interpretation of DMD results for the transient from a stationary state to a post-transient attractor.
Therefore, extracting the transient dynamics from flow data as in the previous simple example
would provide a complementary perspective.
Such an analysis would show how the system changes through intermediate states,
helping us to understand the connection between the initial transient and the final attractor.
We present the results for the cases with the sparsity parameter  $\alpha_{1}=0.05$,\ $0.10$,\ $0.15$, and $0.20$, while fixing the smoothness parameter $\alpha_{2}=10$ and the truncation rank $r=24$.

First, we apply the exact DMD algorithm to the dataset and obtain the DMD modes and eigenvalues, $\{(\bphi_{j}, \lambda_{j})\}$.
Figure~\ref{Fig:Exact DMD Result}(a) shows the real part of the DMD modes (labeled with continuous-time frequencies $\Omega_j = \omega_j/\Delta t$), and the corresponding continuous-time eigenvalues $\mu_j = \log(\lambda_j)/\Delta t$ are plotted in the complex plane in Fig.~\ref{Fig:Exact DMD Result}(b).
Since the DMD eigenvalues and modes arise in symmetric, complex-conjugate pairs, we focus only on the odd modes.
For example, the DMD mode $\bphi_{11}$ has $\Omega_{11}=3.5716$ and represents the vortex shedding, and the DMD mode $\bphi_{7}$ has $\Omega_{7}=7.023$ and represents the second harmonics.
The DMD mode $\bphi_{19}$ exhibits slow dynamics, capturing slow variations in the base flow.

\begin{figure*}[t]
    \centering
    \includegraphics[scale=.7]{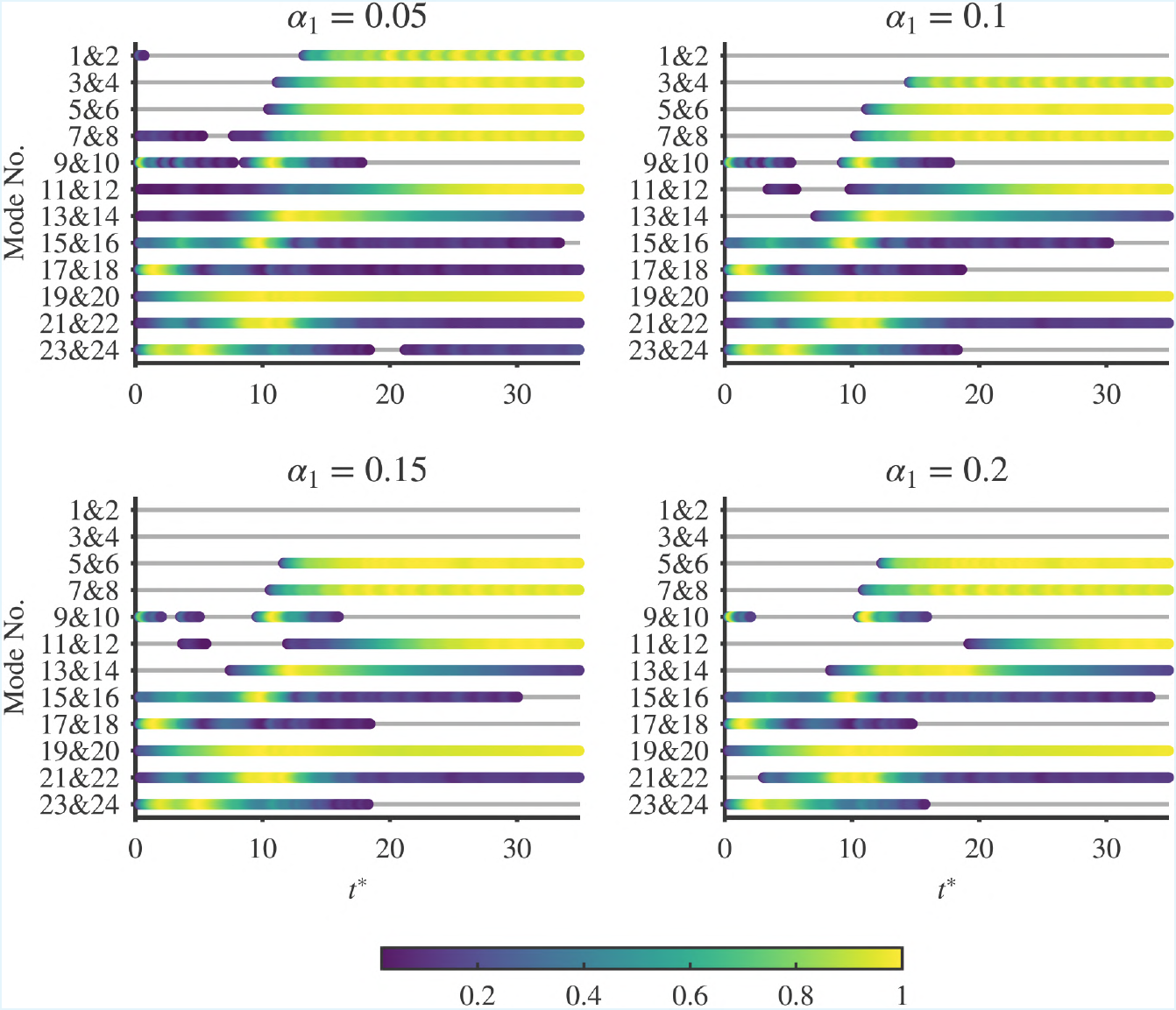}
    \caption{
    {%\bf
    Mode activation diagram.
    Plot of the set $\mathcal{R}_m$ with $|z_{jm}| > 0$ vs convective time $t^{*}$ for each DMD mode $j$ 
    with different values of $\alpha_{1}$ ($= 0.05$, $0.10$, $0.15$, $0.20$) and fixed $\alpha_{2}=10$. 
    In each figure, the bar shows the ``on'' state, whose color represents the value of $|z_{jm}|$ normalized to the range $[0, 1]$, and gray lines indicate the ``off'' state.
    Each pair of even and odd modes ($2n-1$ and $2n$ for integer $n$) are symmetric and exhibit the same activation dynamics.
    A mode activation diagram. It plots the active (on) or inactive (off) state of each DMD mode along the time axis, where the ``on'' state is represented by a color bar indicating the magnitude of the amplitude, and the ``off'' state is indicated by gray lines.
    }}
    \label{Fig:Diagram}
\end{figure*}

We focus on the eigenvalues relatively close to the unit circle on the complex plane. For these modes, the real parts $\mathfrak{R}[\mu_{j}]$ are close to zero, indicating that all modes are slowly growing or decaying.
In the standard DMD including sparsity-promoting DMD, the decomposition is given in the form of Eq.~\eqref{Eq:Decomposition} with constant amplitudes; hence, we can only infer that these modes exhibit slow growth or decay over time. 
While all $r = 24$ modes are included in the optimization problem, we exclude the modes $j = 1$, $19$, and $17$ from our modes of interest for extracting transient activities because their eigenvalues are far from the unit circle, indicating rapid decay that is inconsistent with our slow-amplitude assumption.
As confirmed in the mode activation diagram (see Fig.~\ref{Fig:Diagram}), when appropriate value of $\alpha_{1}$ is chosen, the sparsity constraint appropriately drives the amplitudes of these modes to zero, which is consistent with this expectation.

Figure~\ref{Fig:TemporalStructure} shows the temporal evolution of the amplitudes $\{ z_{jm} \}$ of the DMD modes shown in Fig.~\ref{Fig:Exact DMD Result}(a) extracted by the proposed method.
Each line represents the real part of $z_{jm}e^{i\omega_{j}m}$ in Eq.~\eqref{Eq:Proposed Model}, where $z_{jm}$ is obtained from Eq.~\eqref{Eq:Opt Problem For z modified}. 
The blue, orange, red, and green lines represent the results for $\alpha_{1} = 0.05$, $0.10$, $0.15$, and $0.20$, respectively.
We can observe that the amplitude of each DMD mode now captures the transient or steady dynamics; some modes are active only in the initial transient and vanish in the steady state, indicating that they contribute to the dynamics only in the transient, while some modes are constantly active in the steady state, contributing to the limit-cycle oscillations.

\begin{figure*}[t]
    \centering
    \includegraphics[scale=.65]{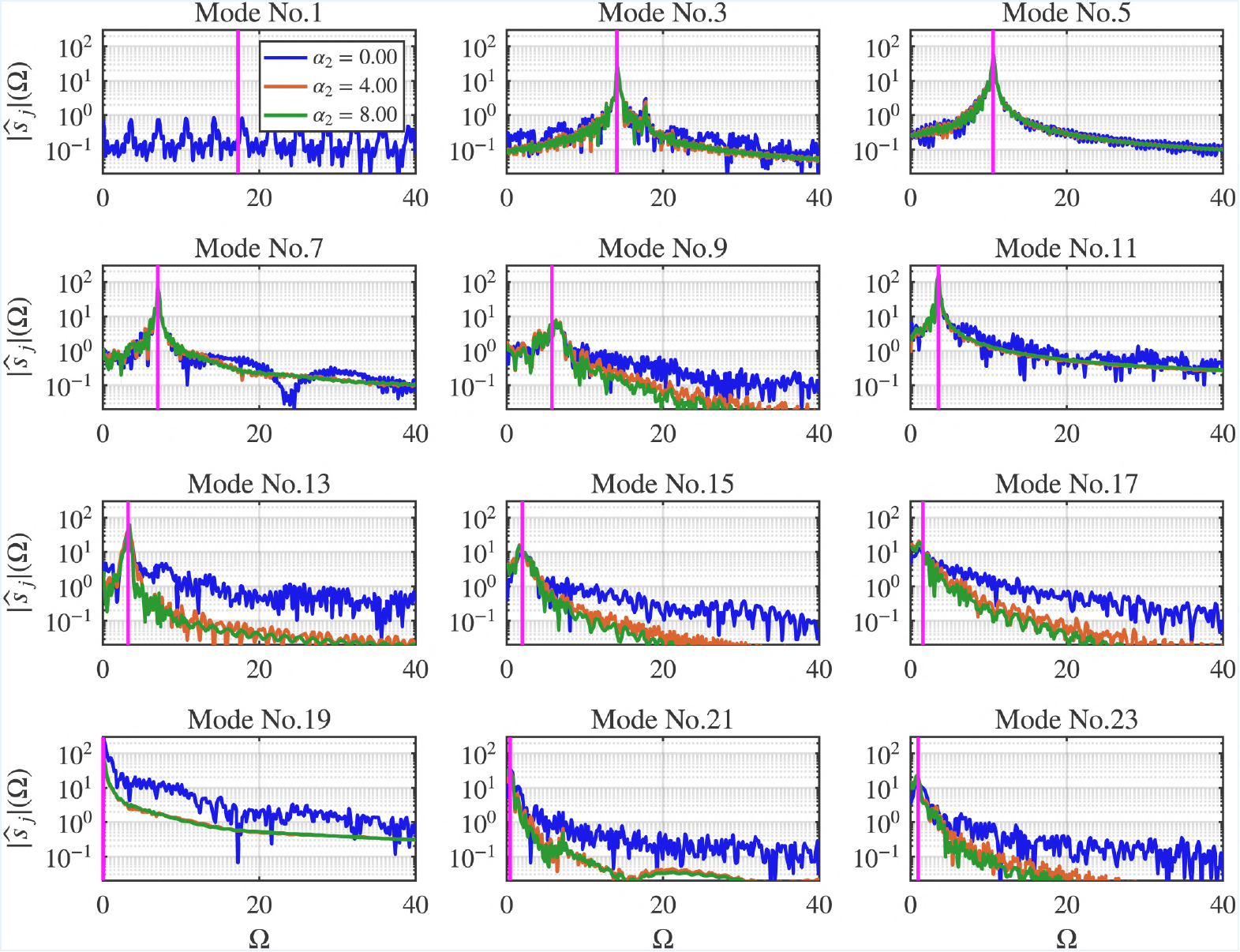}
    \caption{
    {%\bf
    Power spectra of the coefficients $\{ z_{jm}e^{i\omega_{j}m} \}$, plotted against continuous-time frequency $\Omega = \omega/\Delta t$.
    The blue, orange, and green lines correspond to $\alpha_{2}=0.0$, $4.0$, and $8.0$, respectively, with all cases using $\alpha_{1} = 0.10$. 
    In each figure, the magenta line indicates the eigenfrequency $\Omega_{j}$($j=1, 3, 5, 7, 9, 11, 13, 15, 17, 19, 21$, and $23$).
    Graphs showing the power spectra of the extracted DMD mode coefficients. It compares the frequency distributions under three different smoothness parameter settings (blue, orange, and green lines) and shows the spectral spread around the eigenfrequencies (magenta lines).}
    } 
    \label{Fig:FrequencyStructure}
\end{figure*}

Figure~\ref{Fig:ProjectionError} shows the projection error $\epsilon_{m}$ at each time step $m$ for each value of $\alpha_{1}$. The projection error is defined as
\begin{align}
    \epsilon_{m} = 100\left\| \bx_{m} - \bPhi\mathrm{diag}(\bz_{m})
        \begin{bmatrix}
                e^{i m \omega_{1}} \\ \vdots \\  e^{i m \omega_{r}}                               
        \end{bmatrix}
        \right\|_{2}\Big/||\bx_{m}||_{2},
        \label{Eq:ProjectionError}
\end{align}
where $\bz_{m}$ is the solution of Eq.~(\ref{Eq:Opt Problem For z modified}). 
The error generally decays with the time step $m$, indicating that fewer modes are active in the later stage. 
As the sparsity parameter $\alpha_{1}$ increases, the error increases due to the stronger sparsity constraint, but the representation becomes more interpretable because fewer modes are activated at each time step $m$ (see Fig.~\ref{Fig:Diagram}). 

In the late stage ($t^{*}\in[20, 35]$), where the system is in the limit-cycle regime, the projection error $\epsilon_{m}$ remains within approximately $10$-$20$\%.
In contrast, in the time interval $t^{*}\in[1, 4]$ of the initial stage, a significant increase in the projection error is observed, reaching up to 50\%.
This large error reflects the inherently nonlinear dynamics in the early transient that cannot be fully captured by the exact DMD modes extracted from the full trajectory.

Figure~\ref{Fig:Diagram} visualizes the temporal domains where each mode is active, characterized by $\mathcal{R}_m$ for several values of the sparsity parameter $\alpha_1$. 
The color indicates the normalized magnitude of $|z_{jm}|$ ranging from $0$ to $1$, and gray lines indicate the ``off'' state, i.e., $|z_{jm}| = 0$. 
By comparing the results for different values of $\alpha_1$, we can observe how the sparse regularization influences the temporal activity of the DMD modes.
As $\alpha_{1}$ increases, the number of active modes at each $m$ decreases and the switching of modes becomes more distinct, i.e., they become more interpretable. 
However, the projection error also increases, namely, there is a trade-off between the interpretability and reconstruction accuracy.

The mode activation diagram (Fig.~\ref{Fig:Diagram}) reveals a physically coherent picture of the laminar-wake transition. 
Below, we describe its main stages using the $\alpha_1=0.2$ case as a representative example.

\subsubsection{Initial to intermediate stage ($t^{*}\lesssim15$)}
Multiple modes become active in rapid transition as the wake asymmetry grows and alternating vortex shedding emerges.
In particular, especially for modes $j=9$\&$10$, $j=15$\&$16$, $j=19$\&$20$, and $j=21$\&$22$, the activation intensity reaches its highest values around $t^{*}\approx 10$.
This peak corresponds to the moment when the previously near-symmetric flow structure breaks down and rolls up into the first concentrated vortex pair, as visualized in the snapshots of Fig.~\ref{Fig:Simulation Result}~(a2).
As the symmetry is lost, the energy initially concentrated in the steady and symmetric recirculation zone is rapidly redistributed to sustain the developing oscillatory components and the new wake structure~\cite{ROM,Koopman episodic memory}.

This structural transition is captured through two groups of modes in the proposed method:
(i)~Low-frequency modes $j=19\&20$ ($\Omega_{19}=0.037$) and $j=21\&22$ ($\Omega_{21}=0.43$) represent a large-scale modification of the flow underlying flow structure.
These modes account for the global movement and deformation of the fluid as it transitions from a quiescent state to a periodic shedding regime.
(ii)~Modes such as $j=9\&10$ and $j=15\&16$ are activated with maximum intensity during this stage. 
These modes do not match with the final K\'{a}rm\'{a}n vortex shedding frequency.
Their diverse combination is essential to capture the highly non-stationary and nonlinear process of early wake development.

\subsubsection{Late stage ($t^{*}\gtrsim15$)}
Once the transient modes decay, the flow reaches a stable periodic limit-cycle attractor, as confirmed by the Lissajous plot in Fig.~\ref{Fig:Simulation Result}~(c).
On this attractor, the dynamics are governed by a small set of modes that constitute the nonlinear K\'arm\'an vortex street.
The dominant shedding mode $j=11\&12$ ($\Omega_{11}=3.57$) represents the fundamental frequency and serves as the primary component of the periodic wake. 
Its harmonics, $j=7\&8$ ($\Omega_7 \approx 2\Omega_{11}$) and $j=5\&6$ ($\Omega_5 \approx 3\Omega_{11}$), are also active. 
The simultaneous activation of the fundamental frequency and its harmonics indicates that the flow has reached a fully nonlinear limit cycle.
The base-flow mode ($j=19$\&$20$) remains active throughout, which is consistent with the slowly varying mean flow.
In contrast, the transient modes from the intermediate stage ($j=9\&10, 17\&18, 23\&24$) vanish as $t^{*}$ passed $15$, confirming that they played a role only in approaching to the attractor.

In the mode activation diagram, a counterintuitive result that fewer active modes are active during the highly nonlinear transient stage than in the periodic limit-cycle stage emerges. 
This result reflects the inherent limitation of the linear DMD framework and the sparsity and smoothness constraints.
During the early transient, the flow undergoes rapid reorganization that is fundamentally nonlinear and cannot be fully captured by the DMD modes and the frequencies.
Rather than overfitting to these inherently uncapturable dynamics, which would reduce the reconstruction error at the cost of physically uninterpretable mode assignments, the sparsity and smoothness constraints in our method avoid such overfitting by selecting only the modes that yield an interpretable representation consistent with the slow-amplitude assumption in Eq.~\eqref{Eq:Opt Problem For z}.
From this perspective, this counterintuitive result serves as evidence of the strongly nonlinear nature of the early transient dynamics.
The reconstruction error in this period (Fig.~\ref{Fig:ProjectionError}) is also
consistent with this interpretation.

We can also discuss the results from a frequency perspective.
Figure~\ref{Fig:FrequencyStructure} shows the power spectra of the coefficients $s_{jm} = z_{jm}e^{i\omega_j m}$ of each DMD mode, plotted against the continuous-time frequency $\Omega = \omega/\Delta t$.
In each panel, the magenta vertical line represents the eigenfrequency $\Omega_j$, and the blue, orange, and green lines correspond to $\alpha_2 = 0.0$, $4.0$, and $8.0$, respectively (with $\alpha_1 = 0.10$ fixed throughout).
Without the smoothness constraint ($\alpha_2 = 0$, blue), the spectra are broad, indicating that the extracted coefficient signals contain large high-frequency fluctuations and lack spectral compactness.
For $\alpha_2 = 8.0$ (green), the spectrum is sharply localized at $\Omega_j$, producing IMF-like coefficients.
This indicates that the smoothness constraint localizes the frequency components of the temporal evolution of each DMD mode, allowing physically interpretable
IMF-like signals to be extracted.

The primary purpose of the proposed method is not simply to minimize the reconstruction error but to identify  the small set of transiently active modes that provide an interpretable representation of the dynamics, i.e., modes whose coefficients can be regarded as simple pure tones.
As shown in Fig~\ref{Fig:Diagram}, the mode activation diagram achieve this objective even when reconstruction error is relatively high.

%%%%%%%%%%%%%%%%%%%%%%%%%%%%%%%%%%%%%%%%%%%%%%%%%%%%%%%%
%%% Discussion and conclusions %%%%%%%%%%%%%%%%%%%%%%%%%
%%%%%%%%%%%%%%%%%%%%%%%%%%%%%%%%%%%%%%%%%%%%%%%%%%%%%%%%

\section{Discussion}

\begin{figure*}[t]
    \centering
    \includegraphics[scale=.7]{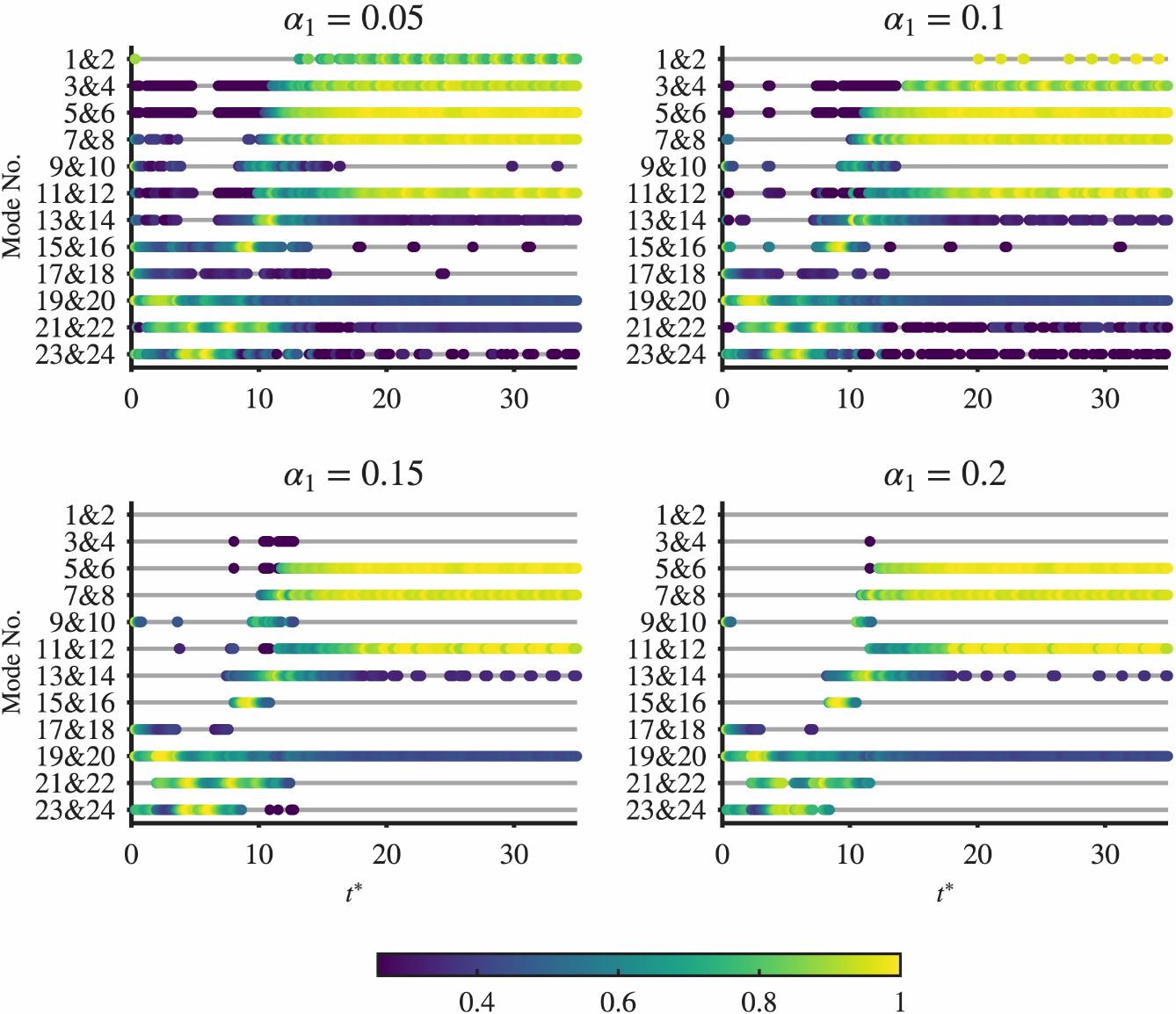}
    \caption{
    Mode activation diagram obtained by the simple
    LASSO-type projection in Eq.~\eqref{Eq:SimpleLasso} with different values of $\alpha_{1}$($= 0.05$, $0.10$, $0.15$, $0.20$), applied to the same dataset as in Sec.~\ref{Sec:NumericalExperiment}.
    The color indicates the normalized magnitude of $|z_{jm}|$ in the range $[0,1]$, and gray lines indicate the ``off'' state, i.e., $|z_{jm}|=0$.
    A mode activation diagram obtained using a simple LASSO-type projection. It shows the active state (color bar) and inactive state (gray lines) of each mode along the time axis.}
    \label{Fig:SimpleLasso}
\end{figure*}

Here, we discuss limitation of the proposed method, comparison with other related methods, and other relevant issues.

\subsection{Applicability to different types of dynamics}
The framework presented in this study is not restricted to analyzing initial transients before converging to limit cycles.
It also applies to systems under the effect of occasional external disturbances.
When a stable system, such as a limit-cycle oscillator, is disturbed, it temporarily deviates from its steady behavior before eventually returning to its original state.
The sparsity constraint in the proposed method would adaptively identify the onset and decay of these transiently excited modes, providing a data-driven characterization of the recovery process.
It is, therefore, suitable for the analyzing, for example, flow disturbed by a gust as it recovers to the limit-cycle vortex shedding~\cite{Gust}.
We also expect that the proposed method is useful in analyzing systems with low-dimensional chaotic attractors.
In such a system, the proposed method could identify changes between different dynamical regimes, providing a time-resolved visualization of the transitions.

\subsection{Limitation on the number of modes}
The fundamental sparsity assumption of the proposed method is justified for data with a relatively small number of dominant oscillatory structures.
Thus, for fluid flows, it is most suited for the laminar limit-cycle wakes studied in Secs.~\ref{Sec:SimpleExample} and \ref{Sec:NumericalExperiment}.
For spatiotemporal turbulent transients, the energy is distributed across a broad continuous spectrum,~\cite{turbulent flows} and no small set of modes can be considered sparse at any given instant.
Extending the proposed framework to such flows would require richer mode bases (e.g., from KDMD~\cite{KDMD}) combined with more flexible sparsity priors, rather than strict ``on/off,'' mode activation.
This remains an important direction for future work.\\

\subsection{Role of sparsity with non-orthogonal modes}
The proposed method relies on the DMD modes as the starting point.
Since DMD modes are generally not spatially orthogonal, simply setting $z_{jm}=0$ may not truly remove the spatial component of mode $j$.
Indeed, projecting the remaining active modes onto the full space may still produce a component in the direction of the deactivated mode.
However, the purpose of introducing sparsity in our method is to improve interpretability rather than to strictly eliminate certain components. 
At each time step, the sparsity penalty identifies which modes most parsimoniously account for the observed data. 
This interpretability is most beneficial when the modes are well separated in frequency, which is guaranteed by the smoothness constraint (Sec.~\ref{Sec:FrequencyPerspective}).

\subsection{Comparison with the simple projection method}
The simplest and most straightforward alternative to our method would be to apply a LASSO-type sparse regression to the DMD modes at each time step,
\begin{equation}
    \bz_{m} = \underset{\by\in\mathbb{C}^{r}}{\mathrm{argmin}}\|\bx_{m}-\bPhi\by\|^2_{2} + \alpha_{1}\|\by\|_{1}, 
    \label{Eq:SimpleLasso}
\end{equation}
which reduces to $\bz_{m}=\bPhi^{\dag}\bx_{m}$ when 
$\alpha_{1}=0$.
This approach simply seeks a least-squares fit of each snapshot into the DMD modes independently at each time step.
Figure~\ref{Fig:SimpleLasso} shows the mode activation diagram
obtained by applying Eq.~\eqref{Eq:SimpleLasso} with $\alpha_{1}=0.05$, $0.1$, $0.15$, and $0.2$
to the same dataset as in Sec.~\ref{Sec:NumericalExperiment}.
In contrast to the proposed method (Fig.~\ref{Fig:Diagram}), the resulting active coefficient sets can rapidly and inconsistently switch between different subsets of modes across time steps.
As a result, it is difficult to identify which modes are dynamically active at any given instant.
Furthermore, this formulation provides no mechanism to control the spectral localization of the extracted components. 
The active modes are selected solely based on instantaneous fit quality, without any preference for frequency or temporally persistent structures.
By contrast, our proposed method focuses on oscillatory transients and recovers the transient localization in a more consistent manner as demonstrated in Sec.~\ref{Sec:NumericalExperiment}.

\subsection{Comparison with windowed and online DMD}
Our approach is fundamentally different from the windowed DMD~\cite{Window DMD1,Window DMD2} and online DMD methods.
Those approaches re-estimate the DMD modes themselves within each time window, so different windows may yield different, possibly incomparable spatial structures.
By contrast, the proposed method fixes a single global set of DMD modes extracted from the full trajectory and tracks only the time-varying amplitudes.
This makes it possible to consistently pinpoint at which time does a specific vortex-shedding frequency first becomes active.
The proposed method is, therefore, considered to be particularly suited for physically interpretable temporal labeling of dynamical events (e.g., onset of vortex shedding, mode switching) across the whole trajectory.

\subsection{Possible generalization using recursive DMD}
Combining the present method with the recursive DMD (RDMD)~\cite{RDMD}, whose modes are orthogonal and retain pure frequencies, could further sharpen this interpretability.
The proposed method requires only the snapshot data $\{\bx_{m}\}_{m=0}^{M}$ and the set of DMD modes and frequencies, $\{(\bphi_{j}, \omega_{j} = \arg(\lambda_{j}))\}_{j=1}^{r}$ and can, therefore, be seamlessly integrated with other types of DMD-type methods without modifying their frameworks.
One notable example is the recursive dynamic mode decomposition (RDMD)~\cite{RDMD}, which was developed to analyze nonlinear transient dynamics.
In RDMD, the resulting modes are orthogonal and retain pure frequencies. Thus, combining RDMD with the present method may lead to more interpretable and distinguishable temporal activations of the extracted modes.
Furthermore, by employing kernel DMD~\cite{KDMD} or extended DMD~\cite{EDMD}, the proposed framework can naturally incorporate nonlinear observables. 
It allows for a richer representation of nonlinear complex dynamics and enables improved characterization of transient phenomena not adequately captured by linear observables alone.

\section{Conclusions}

In this paper, we proposed a simple extension of DMD for analyzing transient oscillatory dynamics in high-dimensional time-series data. 
By introducing time-varying amplitudes for DMD modes and employing both sparsity and smoothness regularization, our method could
adaptively identify which modes and their corresponding frequencies are active at each moment of time and when they contribute to the system's evolution. 
We have illustrated the utility of this method through an example of simple spatiotemporal signal and 
then applied it to a
transient fluid flow.
As demonstrated in the mode activation diagram, 
the proposed method can successfully extract oscillatory transient behaviors in a physically interpretable manner.

Our simple extension to the DMD method provides an alternative viewpoint on the systems exhibiting transient phenomena, which can be useful in analyzing complex, non-stationary data of various origins.
In future work, several directions need to be explored to further enhance the proposed framework.
First, the selection of regularization parameters 
could be automated using cross-validation or information criteria to balance the reconstruction accuracy and interpretability. 
Second, further theoretical analysis of the identifiability and uniqueness of the extracted time-varying amplitudes would provide deeper insights into the method's capabilities and limitations.

\section*{Acknowledgments}
We thank K. Taira and Y. Kawahara for helpful comments.
We are also thankful to the reviewers for detailed comments, which significantly helped improving the presentation of this paper.
We acknowledge JSPS KAKENHI (Nos.~25H01468, 25K03081, and 22H00516) for financial support.

\section*{Data Availability}
The data that support the findings of this study are available within the article.
The source codes used for generating the data and figures in this paper are openly available in GitHub at \url{https://github.com/Tanaka-Yutaroo/sparse-smooth-dmd}, Ref.~\cite{Code}.

\end{document}